\documentclass[fleqn,usenatbib]{mnras}

\usepackage[T1]{fontenc}
\usepackage{xcolor}
\DeclareRobustCommand{\VAN}[3]{#2}
\let\VANthebibliography\thebibliography
\def\thebibliography{\DeclareRobustCommand{\VAN}[3]{##3}\VANthebibliography}

\newcommand{\specialcell}[2][c]{%
  \begin{tabular}[#1]{@{}c@{}}#2\end{tabular}}

\usepackage[normalem]{ulem}
\newcommand{\rev}[2]{#2}

\usepackage{graphicx}	% Including figure files
\usepackage{amsmath}	% Advanced maths commands
\usepackage{amssymb}	% Extra maths symbols

\def\gaia{\textit{Gaia}}
\def\kms{km s$^{-1}$}
\def\kmskpc{km s$^{-1}$ kpc$^{-1}$}

\defcitealias{ramos18}{R18}

\title[Moving groups across radius]{Moving groups across Galactocentric radius with Gaia DR3}

\author[Lucchini, et al.]{
	Scott Lucchini$^{1\dagger}$, Emil Pellett$^1$, Elena D'Onghia$^{1,2}$, J. Alfonso L. Aguerri$^{3,4}$ 
	\\
	$^{1}$Department of Physics, University of Wisconsin - Madison, Madison, WI, USA \\
	$^{2}$Department of Astronomy, University of Wisconsin - Madison, Madison, WI, USA \\
	$^{3}$Instituto de Astrofísica de Canarias; C/ Vía Láctea s/n, 38200, La Laguna, Spain\\
    $^{4}$Departamento de Astrofísica, Universidad de La Laguna, E-38206 La Laguna, Spain \\
    $^\dagger$lucchini@wisc.edu
}

\date{Accepted 2022 November 24. Received 2022 November 18; in original form 2022 June 21}

\pubyear{2022}

\usepackage{newtxtext,newtxmath}
\usepackage{cancel}
\begin{document}
	\label{firstpage}
	\pagerange{\pageref{firstpage}--\pageref{lastpage}}
	\maketitle

\begin{abstract}
The kinematic plane %(azimuthal vs radial velocity) 
of stars near the Sun has proven an indispensable tool for untangling the complexities of the structure of our Milky Way (MW). With ever improving data, numerous kinematic ``moving groups'' of stars have been better characterized and new ones continue to be discovered. Here we present an improved method for detecting these groups using \textit{MGwave}, a new open-source 2D wavelet transformation code that we have developed. %in Python, \textit{MGwave}. 
Our code implements similar techniques to previous wavelet software; however, we include a more robust significance methodology and also allow for the investigation of underdensities which can eventually provide further information about the MW's non-axisymmetric features. Applying \textit{MGwave} to the latest data release from \textit{Gaia} (DR3), we detect \rev{45}{47} groups of stars with coherent velocities. We reproduce the majority of the previously detected moving groups in addition to identifying \rev{four}{three} additional significant candidates: \rev{one near Antoja12-GCSIII-13, one near GMG 5, and two with very low $V_R$ and similar $V_\phi$ to Hercules}{one within Arcturus, and two in regions without much substructure at low $V_R$}. Finally, we have followed these associations of stars beyond the solar neighborhood, from Galactocentric radius of 6.5 to 10~kpc. Most detected groups are extended throughout radius indicating that they are streams of stars possibly due to non-axisymmetric features of the MW.
% We also find several groups that are isolated in radius suggesting a different origin.
% We additionally find that Hercules becomes more prominent at smaller radii and in azimuth towards the semi-minor axis of the MW bar. This is an independent confirmation that the MW has a long bar with Hercules stars being at the corotation resonance with the bar, trapped at the L4/L5 Lagrange points.
%which could mean they are formed through different methods. In future work, we plan to explore the chemistry of these groups and combined with our wavelet transformation we will be able to continue to constrain the structure and dynamics of our Galaxy.
\end{abstract}


\section{Introduction} \label{sec:intro}

Even a relatively small region of the Milky Way (MW) around our Sun contains a wealth of information about the larger properties and non-axisymmetric features of our Galaxy. By studying the motion of nearby stars we can begin to untangle the many complex components of the gravitational potential of the MW including the bar and spiral arms. We have also seen evidence that the stellar disc is out of equilibrium with the discovery of local vertical features like the Radcliffe Wave \citep{alves20} and a more extended vertical kinematic wave \citep{thulasidharan21}.
Since the Hipparcos mission, scientists have slowly been uncovering more and more detail in the kinematic grouping of stars in the solar neighborhood (``moving groups''; % in the solar neighborhood by studying overdensities in the distribution of stars in phase space 
\citealt{eggen96,dehnen98,ramos18}). The intricacies of the azimuthal velocity ($V$; $V_\phi$) vs radial velocity ($U$; $V_R$) distribution of nearby stars shows that the MW is anything but a smooth galactic disk in dynamical equilibrium \citep{dehnen98,antoja18}. By studying the origin of these substructures with the advent of \gaia\ \rev{}{\citep{gaia}} %as our data improves 
and, at the same time, utilizing various theoretical approaches \citep{quillen11,fujii19,monari19,donghia20,trick21,craig21}, we can gain insights %begin to create a map 
on the various components of the MW and understand better galactic structure and evolution.
%we can drastically improve the constraints on our Galaxy's structure and therefore its evolution and future.

\begin{figure*}
    \centering
    \includegraphics[width=\textwidth]{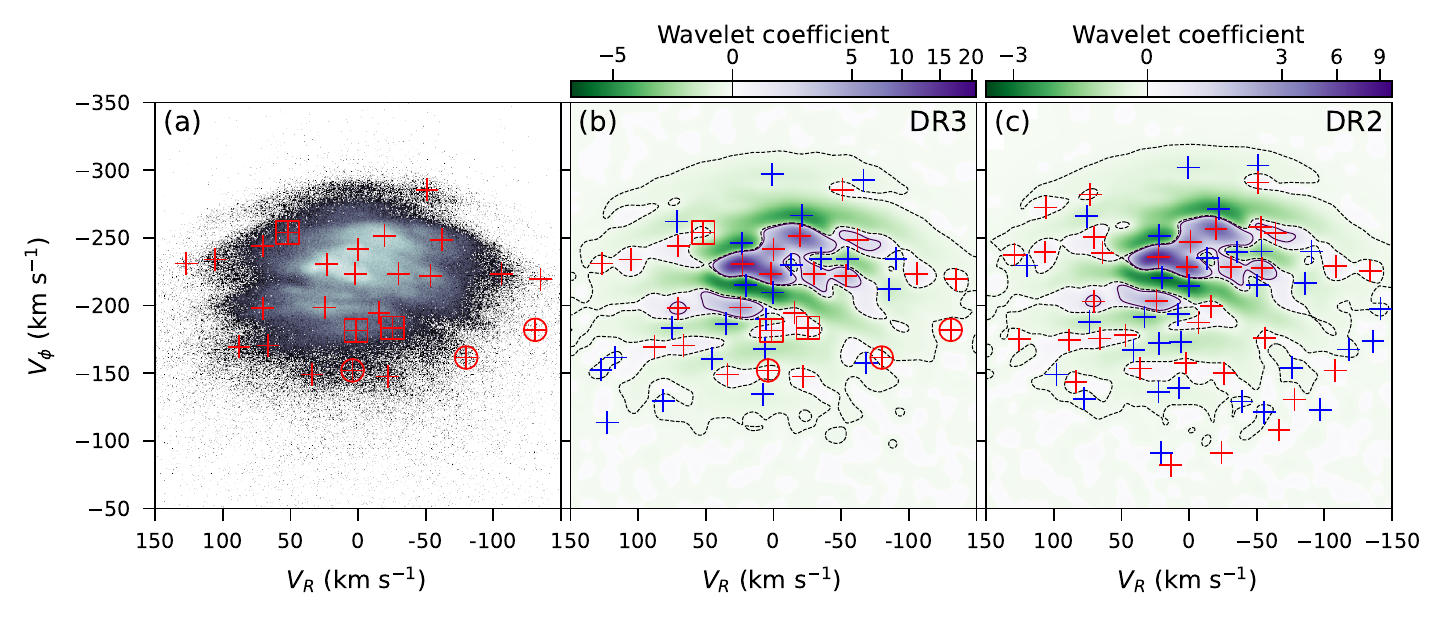}
    \caption{Panel a shows the 2D histogram of all the solar neighborhood stars in the $V_R-V_\phi$ kinematic plane using \gaia\ DR3 with a bin size of 0.5~km~s$^{-1}$. \rev{}{Overplotted are the locations of the wavelet maxima using DR3 data (as in Panel b, see below).} Panels b and c show the results of the wavelet transformation for DR3 and DR2, respectively, shown at a scale of 8-16~km~s$^{-1}$. The purple and green regions depict the relative strength of the positive and negative wavelet coefficients, respectively, \rev{}{normalized by square-root}. Contours are shown at the -0.1\% (dashed) and the 5\% levels. Red markers represent significant overdensities and blue markers represent significant under-densities. Note that only extrema with confidence level $\geq2$ and $P_\mathrm{MC}>0.8$ are shown.
    The DR3 overdensities are also shown overlaid onto panel (a).
    In panels (a) and (b) the \rev{five new group candidates (groups 12, 21, 26, 33, and 34 in Table \ref{tab:dr3})}{three new group candidates (DR3G-25, 26, and 31)} are circled \rev{}{and those groups previously detected but missing from prior studies using \gaia\ are enclosed within squares (Kushniruk17-J4-19, HR1614, and Zhao09-9; marked with asterisks in Table~\ref{tab:dr3})}.
    }
    \label{fig:results}
\end{figure*}

% One of the least well-constrained aspects of the MW's structure is its galactic bar.
One of the least well-constrained features of the MW is its bar. %This elongated stellar overdensity 
This non-axisymmetric feature can have a significant gravitational potential and %due to its non-axisymmetric nature, it 
can affect the distribution of stars through resonances. Depending on the length and pattern speed of the bar, it could provide different explanations for many of the moving groups that we see in the solar neighborhood (SN). For example, previous models of the MW included a short bar with a pattern speed of $\sim$55 \kmskpc where the outer Lindblad resonance (OLR) coincided with the SN \citep{dehnen00,debattista02,monari17}. The OLR creates a bimodality in the kinematic plane of stars around the sun providing one possible explanation for the Hercules group. However, recent observations before \gaia\ DR3 already suggest that the bar is actually long and rotating more slowly ($\sim$40 \kmskpc; \citealt{clarke19,sanders19}). Theoretical models and simulations have shown that Hercules could be formed by stars at the corotation resonance of a long bar \citep{perez-villegas17,donghia20,asano20}. 
Furthermore, several of the significant moving groups in the solar neighborhood are explained by being in resonance with the long bar (the Outer Lindblad resonance (2:1), the 4:1 or the Outer Ultra-Harmonic resonance, and the 6:1 correspond to the Hat, Sirius, and the Horn, respectively; \citealt{monari19}).
% Unfortunately, the current data is not sufficient to break the degeneracy in bar resonances to constrain the pattern speed of the bar. \citep{trick21,trick22}.
The MW's spiral arms have also been shown to have a significant effect on the kinematics of the SN \citep[e.g.][]{antoja09,hunt18,michtchenko18,barros20}.
While looking at the resonances in the solar neighborhood alone are not sufficient to break the degeneracy to discriminate between the long and short bar scenarios \citep{trick21,trick22}, \gaia\ DR3 provided the data to observe the bar in the azimuthal velocity field of the galaxy and indicate that the pattern speed is between 38-42 km/s/kpc \citep{drimmel22}. 
%From these data it seems that the bar is long with a pattern speed of $\sim$40 \kmskpc. 
% As the data improve and we continue to better characterize substructures within the solar neighborhood and throughout the MW's disk, we might be able to identify signatures of resonances and further constrain the properties of the MW's bar and spiral arms.

\gaia\ constitutes the largest and most precise database of positions and velocities of stars in the MW to date, which makes it perfect for this exploration \citep{gaia}. %\gaia\ has previously been used to great success in illuminating substructure in the kinematic planes of the solar neighborhood \citep{ramos18,antoja18,bernet22}. 
Its latest release (Data Release 3; DR3; \citealt{gaiadr3}) provides improved astrometry and errors for 1.4 billion stars based on 34 months of data. For this study, we include approximately 34 million stars centered on the Sun for which \gaia\ provides positions, proper motions, parallaxes, and radial velocities.

While much of the structure in the $V_R-V_\phi$ kinematic diagram is clearly visible as overdensities (e.g. Figure~\ref{fig:results}a), more sophisticated methods are required to quantify the wealth of information. One such technique is the wavelet transformation. Much like a Fourier transformation, the wavelet transform (WT) decomposes data into different components based on a given scale \citep[see][and references therein]{starckbook}. When applied to 2D images, the WT can isolate visual structures of different sizes. This technique has been used on a variety of astrophysical data where it allows for the detection of subtle variations from uniformity, e.g. cosmological large-scale structure \citep{slezak93,einasto11}, galaxy cluster distributions \citep{girardi97,darocha05,darocha08}, and the cosmic microwave background \citep{sanz99,rogers16,hergt17}.

Recent work has also utilized the WT to explore the kinematic space of the MW \citep{antoja08,zhao09,zhao14,zhao15,kushniruk17,ramos18,yang21,bernet22}. \citet{ramos18} (hereafter \citetalias{ramos18}) used \gaia\ DR2 and the WT to detect moving groups in the $V_R-V_\phi$ plane and found many arch features covering the majority of previously known moving groups. In addition to detecting 28 new overdensities, they traced $V_\phi$ of the groups over radius and azimuth to compare with the detected ridges in $V_\phi-R$ space \citep{antoja18}. 
%and the expected curves of constant angular momentum. 
The outcome showed that there are kinematic features indicative of both phase mixing processes as well as resonant trapping due to the MW's non-axisymmetric structures.

Similarly, \citet{bernet22} explored a larger region of the MW disk using \gaia\ eDR3 and the WT combined with a breadth-first search to group detected overdensities together in $(R,\phi,Z,V_R,V_\phi)$ space. The WT they use is one-dimensional and was specifically designed to detect and group overdensities into the arches found in \citetalias{ramos18}. They again explore the variation in $V_\phi$ vs $R$ for each detected group in addition to looking at the distribution of $V_\phi$ along $\phi$ and $z$. By comparing with both slow- and fast-bar models, they find several resonances that overlap with the detected groups/arches, however \gaia\ eDR3 was not extended enough to determine the bar's length and pattern speed in order to remove the degeneracy.
%degeneracy between pattern speeds is still evident.

\begin{figure*}
    \centering
    \includegraphics[width=\textwidth]{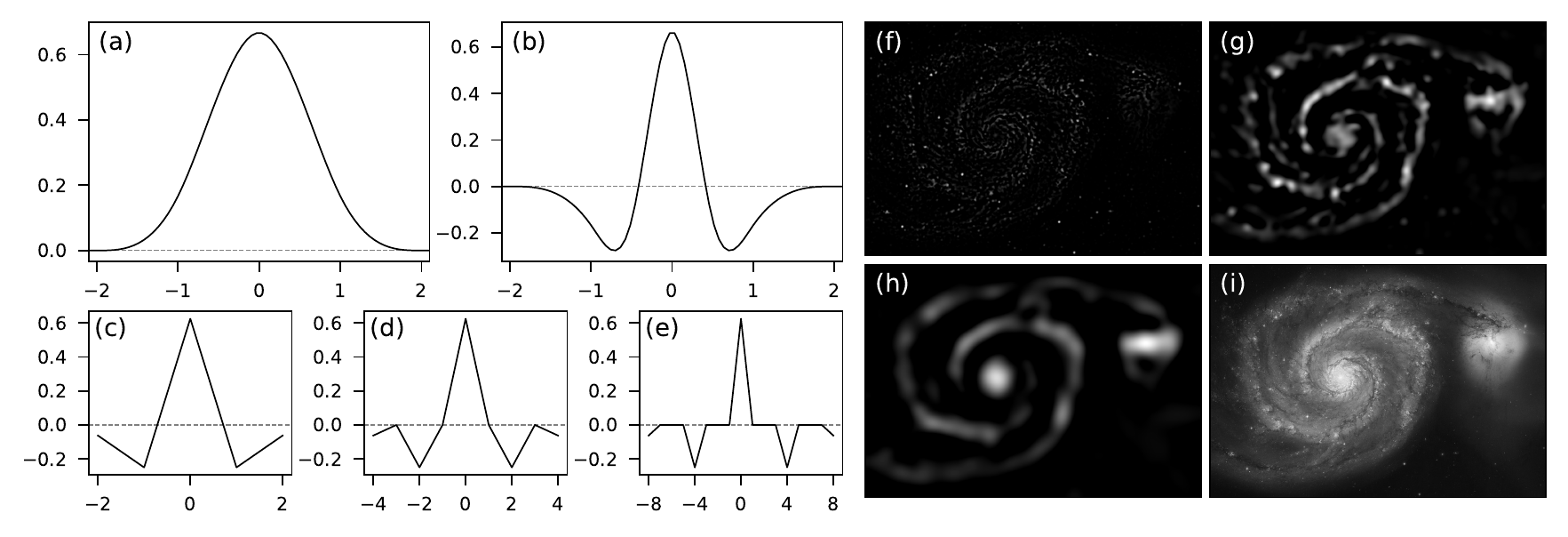}
    \caption{Panels \textbf{a} and \textbf{b} show the continuous B3-spline scaling function, $\phi(x)$, and the corresponding wavelet function, $\psi(x)$, respectively. Panels \textbf{c}$-$\textbf{e} show the discrete wavelet function (generated from the filters $h$ and $g$) for scales $j=0$, $j=1$, and $j=2$. Panels \textbf{f}$-$\textbf{h} show the resulting wavelet transform of the image of M51 shown in panel \textbf{i}.\protect\footnotemark[2]~ The original image is 720 pixels wide by 1037 pixels high and the scales shown in panels \textbf{f}, \textbf{g}, and \textbf{h} are $j=2$, $j=4$, and $j=5$ resulting in the detection of features with sizes of 4-8 pixels, 16-32 pixels, and 32-64 pixels respectively.}
    \label{fig:wavelet_info}
\end{figure*}

For this work, we have developed an open-source WT code for use in Python, \textit{MGwave}\footnote[1]{This code is 
publicly available at \url{https://github.com/DOnghiaGroup/MGwave}.}. %Any bugs can be reported using the ``issues'' tab on the GitHub page.}. 
The code is based on the \'a trous algorithm \citep{starck94,starckbook} and is able to perform the wavelet transformation on any 2D image and output the resultant wavelet coefficients, locations of the extrema, as well as a significance, or confidence level for each extremum (when compared to values resulting from Poisson noise). %In this paper 
We build on previous works by using our %newly developed 
\textit{MGwave} code to analyze the SN as seen by \gaia\ DR3. By performing the full 2D WT, we not only detect new kinematic moving groups, but we are also able to track their extension in the kinematic plane and their location through Galactocentric radius.
% By looking at a larger area within the MW disk, we can get stricter constraints on the potential and properties of the Milky Way.
By identifying moving groups of stars that are extended across the Galactic disk, we can distinguish the large-scale % long-lived 
substructures with a dynamical origin (e.g. those stars that might be in resonance with the bar or spiral arms) from local, transient features.

This paper is structured as follows: Section \ref{sec:methods} outlines our data sample and WT methods, Section \ref{sec:results} shows the main results, Section \ref{sec:disc} discusses the implications of our results within the context of previous works, and conclusions are summarized in Section \ref{sec:conclusions}.

\section{Methods} \label{sec:methods}

\subsection{Gaia Data Sample} \label{sec:data}

% We selected from the approximately 800 million stars that have positions, proper motions, parallaxes, and radial velocities in \gaia\ eDR3\footnote[2]{\gaia\ has not updated the radial velocities since Data Release 2, so in our analysis we use eDR3 positions and proper motions while using radial velocities from Data Release 2}. In order to avoid the known biases caused by inverting the parallax to find distances, we use the geometric distances, along with their errors, computed by \citet{bailer-jones21} \citep{bailer-jones-data}.
We selected from the approximately 34 million stars that have positions, proper motions, parallaxes, and radial velocities in \gaia\ DR3. \rev{We required ``good'' parallaxes with errors less than 20\% ($\omega/\sigma_\omega > 5$) and inverted the parallax values to determine the distances to each star.}{In order to avoid the known biases caused by inverting the parallax to find distances, we use the geometric distances, along with their errors, computed by \citet{bailer-jones21} \citep{bailer-jones-data}.}
% We transformed the six-dimensional \gaia\ observables to Galactocentric cylindrical coordinates with the sun located at $\phi_\odot = 180^\circ$, $Z_\odot = 14$~pc, and $R_\odot = 8.34$~kpc. We take the peculiar motion of the sun with respect to the local standard of rest in cartesian coordinates as $(U,V,W) = (11.1,12.24,7.25)$~km~s$^{-1}$ and the circular velocity of the sun as $V_c = 240$~km~$s^{-1}$ \citep{schonrich10,reid14}.
We transformed the six-dimensional \gaia\ observables to Galactocentric cylindrical coordinates with the sun located at $\phi_\odot = 0^\circ$, $Z_\odot = 5.5$~pc, and $R_\odot = 8.15$~kpc. We take the peculiar motion of the sun with respect to the local standard of rest (LSR) in cartesian coordinates as $(U,V,W) = (10.6,10.7,7.6)$~km~s$^{-1}$ and the circular velocity of the sun as $V_c = 236$~km~$s^{-1}$ \citep{reid19}. $V_R$ is directed out away from the Galactic center and $V_\phi$ is directed against the direction of rotation of the disk (i.e. $\phi$ decreases in the direction of rotation, towards the major axis of the MW bar, and increases counter to the rotation, towards the minor axis of the MW bar).

\rev{Errors on the \gaia\ data were propagated through to Galactocentric cylindrical coordinates through the use of Monte Carlo simulations. We used the \textit{pyia} code \citep{pyia} to run 256 error samples that were transformed using the above process. By taking the standard deviation of the transformed coordinate values, we obtained an estimate of the error in the Galactocentric coordinates and velocities for each star.}{We used Monte Carlo simulations to transform the \gaia\ errors from right ascention, declination, proper motions, and radial velocities (source properties) into the Galactocentric cylindrical coordinates defined above (final properties). Using the \textit{pyia} code \citep{pyia}, we sampled the source \gaia\ data 256 times for each star, assuming a gaussian distribution for each property. By then transforming the sampled properties into Galactocentric coordinates, we could measure the spread in their values (the standard deviation) to determine the errors in the final properties ($R$, $\phi$, $Z$, $V_R$, $V_\phi$, $V_Z$).}
This method does account for correlations between right ascention, declination, \rev{parallax, }{}and proper motion, but does not include correlations for radial velocity \rev{}{or the Bailer-Jones distances.}

% There are 7,180,466 stars with radial velocities and Bailer-Jones distances in \gaia\ eDR3. 
There are \rev{29,947,046}{33,653,049} stars with radial velocities and \rev{$\omega/\sigma_\omega > 5$}{Bailer-Jones distances} in \gaia\ DR3. This is increased by more than a factor of four over eDR3.
% While we hope to explore significantly larger regions of the Galactic disk in the future, we are currently limited by the data. We therefore prioritize the exploration of the ``solar neighborhood'' defined by
We define the ``solar neighborhood'' region as
$|z| < 0.5$~kpc, $-1.5^\circ < \phi < 1.5^\circ$, and $8.05<R<8.25$~kpc which contains \rev{982,879}{997,918} stars. We have also explored additional volumes throughout the Galactic disk by looking at 70 overlapping radial bins of width 0.2~kpc in the range $R=$ (6.4, 10.1)~kpc while maintaining the constraints on $z$ and $\phi$, specifically (6.4, 6.6)~kpc, (6.45, 6.65)~kpc, (6.5, 6.7)~kpc, etc.
% Also, we explored 50 overlapping azimuthal bins of width $3^\circ$ centered on values of $\phi$ ranging from (-25, 25) (while maintaining $8.05<R<8.25$~kpc and $|z|<0.5$~kpc), specifically (-26.5$^\circ$, -23.5$^\circ$), (-25.5$^\circ$, -22.5$^\circ$), etc.
However, we do see a decrease in the number of stars per bin as we reach the limits of this range.

In order to compare with previous works, we have also performed the same analysis with \gaia's Data Release 2 (DR2). We have followed the same procedure as above however we have used the definition of Galactocentric cylindrical coordinates as defined in \citetalias{ramos18} ($Z_\odot=14$~pc, $R_\odot=8.34$~kpc, $(U,V,W) = (11.1,12.24,7.25)$~km~s$^{-1}$, and $V_c = 240$~km~$s^{-1}$ from \citealt{schonrich10} and \citealt{reid14}). We have 
% All the same transformations and cuts in position were performed, except that we 
also required ``good'' parallax values, i.e. $\omega/\sigma_\omega>5$ and distances were calculated by inverting the parallax. This provided us with an identical data set to that of \citetalias{ramos18}. A comparison of our results between DR2 and DR3 is shown in Figure~\ref{fig:results}.

\footnotetext[2]{Credit: NASA, ESA, S. Beckwith (STScI), and The Hubble Heritage Team (STScI/AURA)}

\setcounter{footnote}{2}

\begin{figure*}
    \centering
    \includegraphics[width=0.8\textwidth]{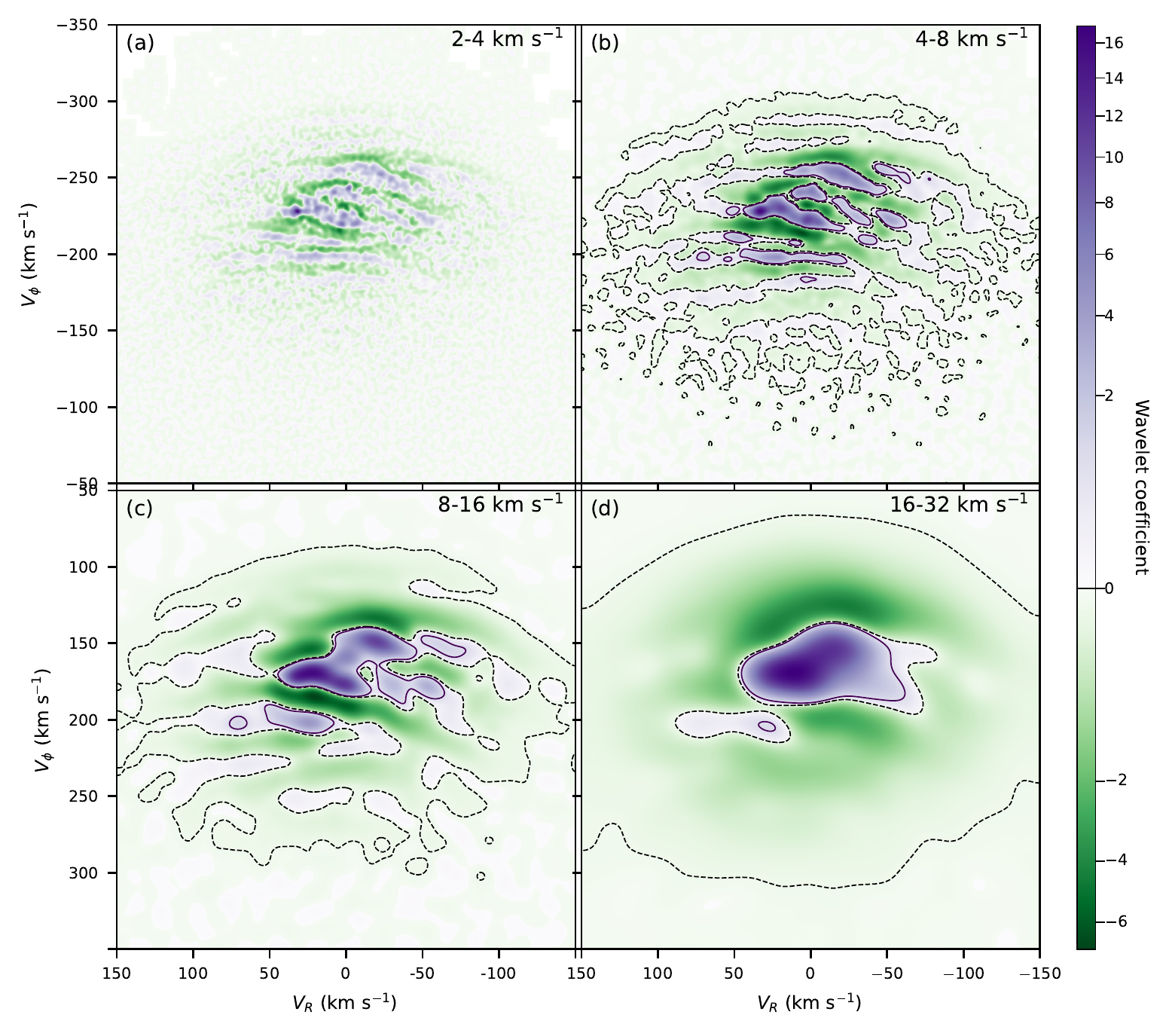}
    \caption{Wavelet coefficient values across the kinematic plane for the solar neighborhood at various wavelet scales using \gaia\ DR3. Panel (a) corresponds to a scale of $j=2$ increasing to a scale of $j=5$ in Panel (d). Given our bin size of 0.5~km~s$^{-1}$, these correspond to physical sizes of 2~km~s$^{-1}$ up to 32~km~s$^{-1}$ (see labels in figure). Contours are shown at the 5\% and -0.1\% (dashed) level (except for Panel (a)).
    % Contours are shown at every 10\% between 10 and 90\% plus 5\% and 99\% (except for Panel (a)). The dashed line contour is at the -1\% level.
    }
    \label{fig:sn_scales}
\end{figure*}

\begin{figure}
    \centering
    \includegraphics[width=\columnwidth]{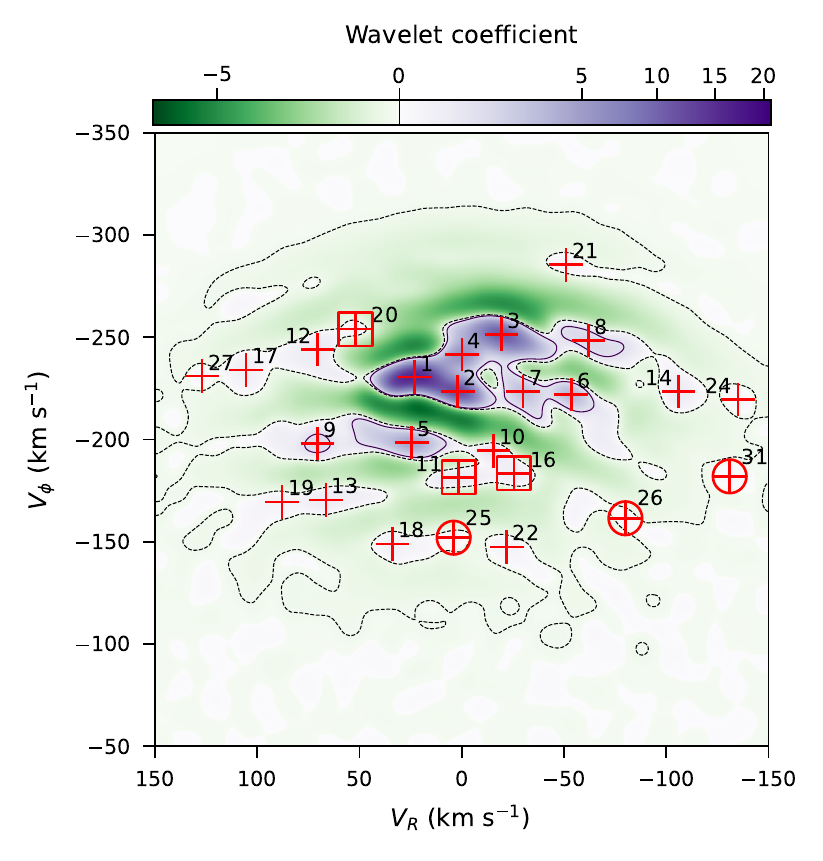}
    \caption{As Figure~\ref{fig:results}b with only maxima shown and numbered by ID as listed in Table~\ref{tab:dr3}. Again, only maxima with confidence level $\geq2$ and $P_\mathrm{MC}>0.8$ are plotted, with circles enclosing the three new group candidates and squares around the three groups not previously seen within \gaia\ data.}
    \label{fig:numbered}
\end{figure}

\subsection{Wavelet Transform Method} \label{sec:wavelet}

% We have developed an open-source wavelet transformation code, \textit{MGwave}\footnote{This code is open source and publicly available at \url{https://github.com/DOnghiaGroup/MGwave}}, based on the \'a trous algorithm \citep{starck94,starckbook}.
Our open-source WT code, \textit{MGwave}, is based on the \'a trous algorithm \citep{starck94,starckbook}.
We have also implemented quantitative analysis to determine the significance of detected structures with respect to Poisson noise \citep{slezak93}. Finally, Monte Carlo simulations are used to propagate data errors through to the wavelet results.

% Our \textit{MGwave} code uses WT techniques to detect overdensities and underdensities in 2D images. We implemented the Starlet transform based on the \'a trous algorithm using a B3-spline scaling function \citep{starckbook06},
Our implementation of the \'a trous algorithm utilizes the Starlet transformation with a B3-spline scaling function \citep{starckbook06},
\begin{multline}
    \phi(x)=B_3(x)=\\
    \frac{1}{12}\left(|x-2|^3-4|x-1|^3+6|x|^3-4|x+1|^3+|x+2|^3\right).
    \label{eq:phi}
\end{multline}
Figures \ref{fig:wavelet_info}a and b shows the continuous scaling function and corresponding wavelet function. Since we are working with pixellated images, we need to discretize these functions. Defined in terms of the $h$ and $g$ filter set \citep{starckbook}, the scaling function corresponds to $h=[\frac{1}{16},\frac{1}{4},\frac{3}{8},\frac{1}{4},\frac{1}{16}]$ and the wavelet function is derived from $g = \delta - h = [-\frac{1}{16},-\frac{1}{4},\frac{5}{8},-\frac{1}{4},-\frac{1}{16}]$ \rev{}{(where $\delta$ is the discretized delta function, i.e. $\delta=[0,0,1,0,0]$)}. These discrete wavelet functions are shown in Figure~\ref{fig:wavelet_info}c$-$e for three different scales ($j=0$, $j=1$, and $j=2$). By applying this separable convolution mask to our image in each dimension sequentially, we obtain the wavelet transformed image (consisting of the values of the wavelet coefficients for each pixel). An example image and its wavelet transforms at three different scales ($j=2$, $j=4$, and $j=5$) are shown in Figure~\ref{fig:wavelet_info}f$-$i. When performing the wavelet transformation at small scales (panel f), the smallest structures in the original image are selected. As we increase the scale of the transformation, larger and larger features are shown. For a more detailed discussion see \citet{starckbook}.

We then use a peak detection algorithm to find local minima and maxima in the wavelet transformed image. We require that detected extrema are separated by at least the wavelet scale size ($2^j$). This is accomplished using the \texttt{peak\_local\_max} function in the \texttt{scikit-image} package \citep{scikit}. We then ensure that if there are two peaks or two troughs within $2^j$ pixels, we only keep the extremum with the larger wavelet coefficient. Once we have located the extrema, we then calculate the significance of each peak and trough to determine whether or not it could be an artifact of Poisson noise.

% When calculating the significance of the wavelet coefficients (see below), we use a method based on the probability density \citep{slezak93}. This method assumes independence between the coefficients. Therefore we limit our extrema to those separated by at least the scale of the wavelet transformation so as not to introduce undue correlations. Additionally, that separation forces detected structures to be indeed detected at the desired scale, essentially ignoring fluctuations due to noise. When two extrema are within this minimum distance, we select the coefficient with the larger value since that pixel is more likely to be the significant feature. That is because the value of the wavelet coefficient directly corresponds to the ``strength'' of the detected structure. Once we have located the extrema, we then calculate the significance of each peak and trough to determine whether or not it could be an artefact of Poisson noise.

\subsection{Significance of Detected Extrema} \label{sec:significance}

Given a wavelet coefficient, its significance must be computed to assess the probability that the detected extremum is ``real''. This will give us a confidence level that the value of a wavelet coefficient (pixel in the transformed image) is not due to random Poisson noise. In order to calculate this, we can integrate the WT probability density function, $p_n(w)$, to determine the likelihood that a random wavelet coefficient due to Poisson noise has a lower value than a wavelet coefficient of value $w$ (\citealt{slezak93}; i.e. larger values of $F(w)$ mean $w$ is more significant):
\begin{align}
    \label{eq1}
	F(w) = \int_{-\infty}^w p_n(x)\;dx
\end{align}

The probability density function depends on both the specific wavelet function chosen (in its continuous form, e.g. Figure~\ref{fig:wavelet_info}b), and also on the number of events used to determine the wavelet coefficient. As stated above, we use a B3-spline as the scaling function, $\phi$ (Equation \ref{eq:phi}). At each wavelet scale, $j$, we dilate the scaling function by a factor of $2^j$ and then renormalize it such that
\begin{align}
\int_{-\infty}^{\infty}\phi\left(\frac{x}{2^j}\right)\;dx = 1.
\end{align}
We then compute the continuous wavelet function (in 2D), $\psi(x,y)$, by looking at the difference between the scaling functions at two successive scales \citep{starckbook}.
\begin{align}
\frac{1}{4}\psi\left(\frac{x}{2},\frac{y}{2}\right) = \phi(x,y) - \frac{1}{4}\phi\left(\frac{x}{2},\frac{y}{2}\right)
\end{align}
where $\phi(x,y)=\phi(x)\phi(y)$.
% See Appendix \ref{a:psidef} for derivation.

The number of events also affects the probability density function. In the case of a 2D histogram (for example, the $V_R-V_\phi$ kinematic plane used later in this paper), the number of events represents the total number of stars within the bins used in calculating the wavelet coefficient. If there is only one event, the probability to get any given wavelet coefficient is represented by the histogram of the wavelet function, $H_1$. For two events, each has the probability represented by the histogram of the wavelet function and since they are independent of each other, we can take the autoconvolution of the histogram to represent the PDF for two events \citep{slezak93}. Therefore for $n$ events, the PDF is $n-1$ autoconvolutions of the PDF for a single event.
\begin{align}
	p_n(x) = H_1 * H_1 * \cdots * H_1
\end{align}
We compute the histogram of the wavelet function using the kernel density estimator (\texttt{kdeplot}) from the \texttt{seaborn} python package \citep{seaborn}.
As described in \citet{slezak93}, a maximum must have $n\geq3$ and a minimum must have $n\geq4$ in order for the significance calculation above to be valid.

%%% This describes the reduction method %%%
% Dilating the continuous wavelet function doesn't affect the histogram of the function, so therefore the PDF above is independent of scale $j$.

% We follow the method outlined in \citet{starck} Section 2.5 to reduce the PDF and the wavelet coefficient for comparison. This modifies the x-coordinate scaling of the PDF by dividing by its standard deviation (which is equal to the square root of the number of event, $n$, multiplied by the standard deviation of the continuous wavelet function, $\psi(x)$):
% \begin{align*}
% 	c = \frac{x}{\sqrt{n}\sigma(\psi)}
% \end{align*}

% We also then need to reduce the calculated wavelet coefficients. Here we continue to follow the method in \citet{starck}:
% \begin{align*}
% 	w^r = \frac{w}{\sqrt{n}\sigma(\psi_j)} = \frac{w}{\sqrt{n}\sigma(\psi)}2^{Dj}
% \end{align*}
% where $D$ is the number of dimensions.

% Now we compute the integral of the PDF to determine the significance of the given wavelet coefficient using equation (\ref{eq1}).
% \begin{align*}
% 	F(w) = \int_{-\infty}^{w}p_n(x)\;dx
% \end{align*}

Therefore, $F(w)$ (Equation~\ref{eq1}) can be used to determine the confidence level of each extremum via thresholding. We followed the method in \citetalias{ramos18} setting confidence levels based on these significance values:
\begin{equation}
\begin{aligned}
	&0: F(w) < \epsilon_{1\sigma} \\
	&1: \epsilon_{1\sigma} \leq F(w) < \epsilon_{2\sigma} \\
	&2: \epsilon_{2\sigma} \leq F(w) < \epsilon_{3\sigma} \\
	&3: F(w) \geq \epsilon_{3\sigma}
\end{aligned}
\end{equation}
where $\epsilon_{n\sigma}$ corresponds to the integral of the normal distribution, $N(0,1)$, from $-\infty$ to $n$. This gives $\epsilon_{1\sigma} \approx 0.841$, $\epsilon_{2\sigma} \approx 0.977$, and $\epsilon_{3\sigma} \approx 0.999$.

Following previous works \citepalias{ramos18}, we consider any extremum to be significant if it has a confidence level $\geq2$.

\begin{table*}
\begin{tabular}{rrrlcrccl}
& \multicolumn{1}{c}{$V_R$} & \multicolumn{1}{c}{$V_\phi$} & Name & CL & $P_\mathrm{MC}$ & Wavelet & Stars & Refs \\
& (\kms) & (\kms) & & & & & & \\\hline
 1 &   23.0 & -230.5 & Hyades             & 3 & 1.00 & 15.9831 & 139,712 & 1,2,3,5,7 \\
 2 &    2.0 & -223.5 & Pleiades           & 3 & 1.00 & 11.9514 & 145,719 & 1,2,3,5,7 \\
 3 &  -19.5 & -251.5 & Sirius             & 3 & 1.00 & 10.6615 & 111,631 & 1,2,3,4,5, \\
 4 &    0.0 & -241.5 & Coma Berenices     & 3 & 1.00 & 5.7824 & 147,895 & 1,2,3,4,5 \\
 5 &   24.5 & -198.5 & Hercules II        & 3 & 1.00 & 4.3941 & 54,221 & 2,3,5,7 \\
 6 &  -53.5 & -222.0 & Dehnen98-14 (Horn) & 3 & 1.00 & 3.1135 & 44,473 & 1,2,3,5 \\
 7 &  -30.0 & -223.5 & Dehnen98-6         & 3 & 1.00 & 2.5743 & 86,079 & 1,2,5 \\
 8 &  -62.0 & -248.5 & $\gamma$Leo        & 3 & 1.00 & 1.3055 & 27,352 & 2,3,5,7 \\
 9 &   70.5 & -198.0 & $\epsilon$Ind      & 3 & 1.00 & 1.2055 & 16,928 & 3,5,7 \\
10 &  -15.5 & -194.5 & Liang17-9          & 3 & 1.00 & 0.6758 & 33,501 & 7 \\
11 &    1.5 & -181.5 & Kushniruk17-J4-19* & 3 & 0.99 & 0.4585 & 26,148 & 6 \\
12 &   70.5 & -244.0 & Antoja12-GCSIII-13 & 3 & 1.00 & 0.3995 & 12,169 & 3 \\
13 &   66.5 & -170.5 & GMG 1              & 3 & 1.00 & 0.3706 & 7,728 & 8 \\
14 & -106.0 & -223.5 & Antoja12-12        & 3 & 1.00 & 0.3537 & 4,289 & 3 \\
15 & \textbf{88.5} & \textbf{-202.0} & \textbf{DR3G-15} & \textbf{3} & \textbf{0.55} & \textbf{0.2564} & \textbf{7,782} & \textbf{This work} \\
16 &  -25.5 & -183.5 & HR1614*            & 3 & 1.00 & 0.2453 & 18,636 & 1,5,7 \\
17 &  105.5 & -234.0 & Antoja12-16        & 3 & 1.00 & 0.2380 & 2,750 & 3 \\
18 &   34.0 & -149.0 & $\eta$Cep          & 3 & 1.00 & 0.1813 & 4,299 & 3,5 \\
19 &   88.0 & -169.5 & GMG 3              & 3 & 0.98 & 0.1229 & 3,701 & 8 \\
20 &   52.0 & -254.0 & Zhao09-9*        & 3 & 1.00 & 0.1210 & 16,252 & 2 \\
21 &  -51.0 & -285.5 & GMG 4              & 3 & 1.00 & 0.0884 & 1,913 & 8 \\
22 &  -22.0 & -147.5 & Antoja12-17        & 3 & 1.00 & 0.0835 & 3,123 & 3 \\
23 & \textbf{-56.0} & \textbf{-166.0} & \textbf{DR3G-23} & \textbf{3} & \textbf{0.53} & \textbf{0.0549} & \textbf{3,649} & \textbf{This work} \\
24 & -135.0 & -219.5 & GMG 7              & 3 & 1.00 & 0.0412 & 563 & 8 \\
25 & \textbf{4.0} & \textbf{-152.0} & \textbf{DR3G-25} & \textbf{2} & \textbf{0.97} & \textbf{0.0359} & \textbf{5,007} & \textbf{This work} \\
26 & \textbf{-80.0} & \textbf{-161.5} & \textbf{DR3G-26} & \textbf{2} & \textbf{1.00} & \textbf{0.0204} & \textbf{1,355} & \textbf{This work} \\
27 &  127.0 & -231.0 & GMG 8              & 3 & 1.00 & 0.0204 & 662 & 8 \\
28 &  -37.0 & -135.5 & Antoja12-19*       & 2 & 0.68 & 0.0203 & 1,445 & 3 \\
29 &  -93.0 & -184.5 & Bobylev16-23*   & 1 & 0.99 & 0.0178 & 2,395 & 5 \\
30 & \textbf{104.0} & \textbf{-199.5} & \textbf{DR3G-30} & \textbf{1} & \textbf{0.72} & \textbf{0.0150} & \textbf{2,897} & \textbf{This work} \\
31 & \textbf{-131.0} & \textbf{-182.0} & \textbf{DR3G-31} & \textbf{3} & \textbf{0.93} & \textbf{0.0150} & \textbf{404} & \textbf{This work} \\
32 &  -75.5 & -124.5 & GMG 13             & 2 & 0.76 & 0.0140 & 463 & 8 \\
33 & \textbf{-65.5} & \textbf{-131.5} & \textbf{DR3G-33} & \textbf{2} & \textbf{0.64} & \textbf{0.0102} & \textbf{696} & \textbf{This work} \\
34 & \textbf{79.0} & \textbf{-141.5} & \textbf{DR3G-34} & \textbf{1} & \textbf{0.69} & \textbf{0.0081} & \textbf{1,258} & \textbf{This work} \\
35 &  119.0 & -197.0 & GMG 20             & 1 & 0.91 & 0.0069 & 983 & 8 \\
36 & \textbf{-83.5} & \textbf{-111.5} & \textbf{DR3G-36} & \textbf{1} & \textbf{0.49} & \textbf{0.0067} & \textbf{266} & \textbf{This work} \\
37 & \textbf{-97.0} & \textbf{-136.0} & \textbf{DR3G-37} & \textbf{1} & \textbf{0.95} & \textbf{0.0056} & \textbf{327} & \textbf{This work} \\
38 & \textbf{139.5} & \textbf{-190.5} & \textbf{DR3G-38} & \textbf{1} & \textbf{0.98} & \textbf{0.0049} & \textbf{321} & \textbf{This work} \\
39 &  -71.5 & -281.0 & GMG 10             & 1 & 0.90 & 0.0048 & 934 & 8 \\
40 &   73.0 & -276.5 & GMG 11             & 0 & 1.00 & 0.0044 & 1,113 & 8 \\
41 & \textbf{-25.5} & \textbf{-99.5} & \textbf{DR3G-41} & \textbf{0} & \textbf{0.68} & \textbf{0.0028} & \textbf{271} & \textbf{This work} \\
42 &  -20.5 &  -90.5 & GMG 16             & 0 & 0.64 & 0.0011 & 202 & 8 \\
43 & \textbf{-86.0} & \textbf{-279.5} & \textbf{DR3G-43} & \textbf{0} & \textbf{0.71} & \textbf{0.0001} & \textbf{375} & \textbf{This work} \\
44 & \textbf{15.0} & \textbf{-117.5} & \textbf{DR3G-44} & \textbf{0} & \textbf{0.99} & \textbf{-0.0004} & \textbf{595} & \textbf{This work} \\
45 & \textbf{-87.0} & \textbf{-227.0} & \textbf{DR3G-45} & \textbf{0} & \textbf{1.00} & \textbf{-0.0011} & \textbf{10,406} & \textbf{This work} \\
46 &   -4.0 & -113.5 & GMG 17             & 0 & 0.75 & -0.0011 & 520 & 8 \\
47 &  -40.5 & -115.5 & GMG 22             & 0 & 0.83 & -0.0019 & 495 & 8
\end{tabular}
\caption{Moving groups detected using our new wavelet transform on \gaia\ DR3 data. The same naming convention as \citetalias{ramos18} is followed (see their Appendix C for more information). Groups marked with an asterisk (*) are those that have been previously discovered but were not present in the wavelet analysis of \citetalias{ramos18}. Bold lines are groups newly discovered in this work. Columns 5-8 list the output of our analysis: CL denotes the confidence level that a given group is not due to Poisson noise (see Section \ref{sec:significance}); $P_\mathrm{MC}$ gives the percentage of Monte Carlo simulations in which the peak appeared when varying the stellar velocities within Gaia errors (see Section \ref{sec:mcs}); Wavelet gives the magnitude of the wavelet coefficient at the peak; Stars lists the number of stars in a region of kinematic space around the peak corresponding to the scale of the wavelet transformation performed (in this case within a circle of radius 16~km~s$^{-1}$).\\
References: (1) \citet{dehnen98}; (2) \citet{zhao09}; (3) \citet{antoja12}; (4) \citet{xia15}; (5) \citet{bobylev16}; (6) \citet{kushniruk17}; (7) \citet{liang17}; (8) \citetalias{ramos18}}
\label{tab:dr3}
\end{table*}

\subsection{Monte Carlo Simulations} \label{sec:mcs}

To account for underlying uncertainty in the data, we use Monte Carlo simulations to propagate errors through the WT. Uncertainty values can be supplied for the $x$ and $y$ coordinates for each object (i.e. the data used to create the histogram on which the WT is performed) and \textit{MGwave} will simulate new data by pulling random values from gaussian distributions. After running this simulation process many times and performing the WT on each new data set, the code then calculates the number of simulations in which a peak is detected within a circle of diameter $2^j$ (the scale of the WT) around the actual peak. The workflow is as follows:
\begin{enumerate}
    \item Obtain new $x$ and $y$ values for each object by sampling a gaussian distribution with the associated errors.
    \item Run the wavelet routine on the new, simulated data obtaining a list of maxima and minima.
    \item For each extremum in the original data, check if there exists an extremum in the simulated data within a circle of diameter of $2^j$.
    \item Repeat $N$ times.
\end{enumerate}

For the work presented in this Article, we supplied uncertainties in $V_R$ and $V_\phi$ propagated from the \gaia\ data individually for each star (see Section \ref{sec:data}) and performed $N=$ 2,000 iterations. Following previous works \citepalias{ramos18}, we consider any extremum to be independent of \gaia\ errors if it is reproduced in $>80\%$ of the Monte Carlo simulations, i.e. $P_\mathrm{MC} > 0.8$. These values are listed in Table~\ref{tab:dr3}.

\begin{figure}
    \centering
    \includegraphics[width=\columnwidth]{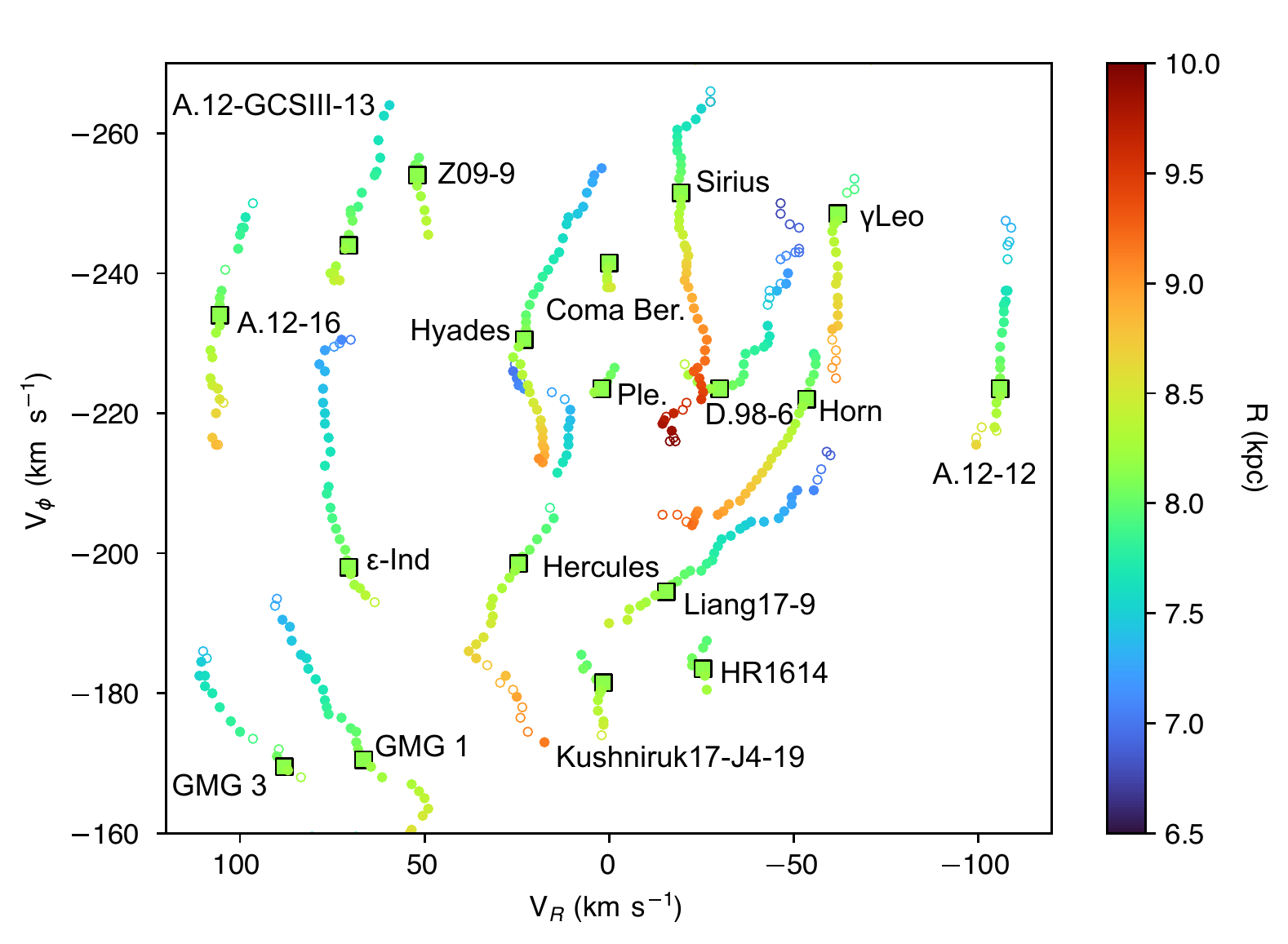}
    \caption{The $V_R$-$V_\phi$ kinematic plane with the locations of detected moving groups shown as a function of radii. The locations of the peaks of the moving groups are shown between radii of 6.5~kpc and 10.0~kpc with radius represented by color. \rev{}{The filled and empty circles represent those peaks with $P_\mathrm{MC}$ $\geq0.8$ and $<0.8$, respectively.} The green square points are the locations of the moving groups in the solar neighborhood region ($8.05<R<8.25$~kpc).}
    \label{fig:worms}
\end{figure}

\section{Results} \label{sec:results}

\rev{}{
Using \gaia\ DR3, we performed the wavelet transformation on the $V_R-V_\phi$ kinematic plane. We first binned the \gaia\ data into 600 bins of size 0.5 \kms\ (in both dimensions; shown in Figure~\ref{fig:results}a). Then we used scales of $j=$ 2, 3, 4, and 5 (shown in Figure~\ref{fig:sn_scales}) for our WT. These scales allow us to detect structures in the histogram with sizes between $\Delta\times2^j$ and $\Delta\times2^{j+1}$ where $\Delta$ is the bin size (0.5 \kms). Since most of the stellar moving group structures have sizes of $\sim10$ \kms, we used the $j=4$ scale for this analysis which corresponds to structures with sizes between 8 and 16 \kms. At smaller scales (Figure~\ref{fig:sn_scales}b) some of the classical moving groups (e.g. Hyades, Coma Berenices, Sirius) break into multiple components, and at larger scales (Figure~\ref{fig:sn_scales}) the groups merge together. While some of the small-scale features are interesting to explore in future works, the goal of this work is to compare with the existing studies of moving groups, so we will focus on the $j=4$ scale below.
}

\subsection{Detected Moving Groups in the Solar Neighborhood} \label{sec:mgs}

\rev{By utilizing \gaia\ DR3, combined with our wavelet code}{From the $j=4$ WT image}, we are able to detect \rev{45}{47} moving groups %(overdensities in the kinematic plane) 
listed in Table~\ref{tab:dr3}. Figure~\ref{fig:results} shows the 2D histogram (Panel~a) as well as the resultant wavelet coefficients and extrema (Panel~b). \rev{}{Both panels show the locations of significant maxima as red crosses, while Panel~b also shows significant minima as blue crosses. The identified moving groups are also shown in Figure~\ref{fig:numbered} which shows only the overdensities with their corresponding ID number (column 1 in Table~\ref{tab:dr3}).} The purple and green shaded regions in Figure~\ref{fig:results}b show the positive and negative wavelet coefficients, respectively. \rev{, with the red and blue crosses marking the locations of the maxima and minima, respectively.}{}
While our results for DR3 are in general very consistent with DR2 (shown in Figure~\ref{fig:results}c), the most significant differences arise from the restriction on the minimum number of stars for a detected maximum. As discussed above, at least 3 stars are required for relative maxima, and 4 stars are required for relative minima in order for consistent significance determination.
% While our results for eDR3 are in general very consistent with DR2 (shown in Figure~\ref{fig:results}b) there are a few differences. The most significant difference is not due to variations in the data between releases, it is due to a change in the criteria for selecting valid overdensities as moving groups. The \textit{MGwave} code implements a minimum number of stars required to count an overdensity as a detection. At least 3 stars are required for relative maxima, and 4 stars are required for relative minima. \citet{slezak93} states that these should be considered effective minimum values for consistent assessment of the significance of each peak.
This cutoff was not implemented in previous works, and for easier comparison with \citetalias{ramos18}, it is disabled in our analysis of the DR2 data below.
% Please reference Section~\ref{sec:prevworks} for a full comparison with previous works.

We are able to detect \rev{16}{15} candidate overdensities in addition to finding \rev{4}{5} previously detected groups that were not detected in \citetalias{ramos18}: Kushniruk17-J4-19 \citep{kushniruk17}, HR1614, \rev{}{Zhao09-9 \citep{zhao09},} Bobylev16-23 \citep{bobylev16}, and Antoja12-19 \citep{antoja12}.
Our \rev{16}{15} candidate groups are numbered with a ``DR3G'' (Data Release 3 Group) prefix in Table~\ref{tab:dr3}. Of our \rev{16}{15} candidate groups discovered, \rev{7}{6} meet the confidence level criteria (CL $\geq2$), 7 meet the Monte Carlo criteria ($P_\mathrm{MC} > 0.8$), and \rev{4}{3} groups meet both criteria (Groups \rev{21, 26, 33, and 34}{25, 26, and 31} in Table \ref{tab:dr3}). \rev{Group 21 is near Antoja12-GCSIII-13 \citep{antoja12} and could indicate further substructure in this region, Group 26 is near the GMG 5 \citepalias{ramos18}, and Groups 33 and 34 are in a region without previous detections at very low $V_R$ with similar $V_\phi$ to Hercules}{Group 25 lies within Arcturus, and Groups 26 and 31 are in regions without much substructure at low $V_R$}. These groups are circled in Figure~\ref{fig:results}a and b.

To compare our wavelet method with previous works, we have reproduced the steps of \citetalias{ramos18}. Following their selection of \gaia\ DR2 data, our code is able to detect all of the top 24 groups listed in their Table~3. We also find 11 of the remaining 20 groups (all of which were new detections not matching any previously known moving group). In addition to the groups found in \citetalias{ramos18}, our wavelet code detects six previously identified groups: Kushniruk17-J5-2, Kushniruk17-J4-19 \citep{kushniruk17}, Dehnen98-11 \citep{dehnen98}, HR1614, \rev{}{Zhao09-9 \citep{zhao09},} Antoja12-19, and Antoja12-15 \citep{antoja12}.

There are also 33 detected overdensities that don't overlap with any of the groups listed in Table 3 or C.1 in \citetalias{ramos18}, however only 3 of these meet the confidence level and Monte Carlo criteria (Groups 30, 39, and 49 in Table~\ref{tab:dr2a}). Groups 39 and 49 use fewer than 3 stars to calculate the wavelet coefficient which is below our cutoff in the DR3 data. Group 30 is detected in the DR3 data as well (Group \rev{30}{25} in Table~\ref{tab:dr3}) \rev{however it has shifted slightly and its confidence level has reduced to 0}{slightly shifted but it remains significant and robust against the Monte Carlo simulations}.

\begin{table}
    \centering
    \begin{tabular}{rlccc}
& Name & \specialcell[c]{$r_\mathrm{min}$\\(kpc)} & \specialcell[c]{$r_\mathrm{max}$\\(kpc)} & \specialcell[c]{Extent\\(kpc)} \\\hline
3 & Sirius & 7.45 & 10.00* & 2.55 \\
5 & Hercules & 6.95 & 9.25 & 2.30 \\
1 & Hyades & 7.20 & 9.05 & 1.85 \\
7 & Dehnen98-6 & 6.75 & 8.35 & 1.60 \\
10 & Liang17-9 & 6.85 & 8.40 & 1.55 \\
6 & Dehnen98-14 (Horn) & 7.85 & 9.30 & 1.45 \\
9 & $\epsilon$Ind & 7.00 & 8.40 & 1.40 \\
14 & Antoja12-12 & 7.30 & 8.65 & 1.35 \\
13 & GMG 1 & 7.25 & 8.55 & 1.30 \\
17 & Antoja12-16 & 7.65 & 8.85 & 1.20 \\
8 & $\gamma$Leo & 7.85 & 9.05 & 1.20 \\
12 & Antoja12-GCSIII-13 & 7.55 & 8.50 & 0.95 \\
19 & GMG 3 & 7.40 & 8.25 & 0.85 \\
11 & Kushniruk17-J4-19 & 7.95 & 8.50 & 0.55 \\
20 & Zhao09-9 & 8.00 & 8.40 & 0.40 \\
4 & Coma Berenices & 8.15 & 8.50 & 0.35 \\
16 & HR1614 & 7.95 & 8.25 & 0.30 \\
2 & Pleades & 8.00 & 8.20 & 0.20
    \end{tabular}\\
    \caption{Radial extent of the moving groups shown in Figure~\ref{fig:worms}. The first column lists the ID corresponding to the groups in Table \ref{tab:dr3}. Note that we found that Sirius extends to the maximum radius of 10~kpc, so it could extend further outwards.}
    \label{tab:rextent}
\end{table}

\subsection{Moving Groups Across the Disk} \label{sec:worms}

One of the most valuable aspects of automated WTs is the ability to quickly and easily detect overdensities and underdensities for an arbitrary dataset. We have used this to analyze different bins of \gaia\ DR3 data to track moving groups throughout Galactocentric radius. We selected 70 radial bins centered on $R$ ranging from 6.5 to 10~kpc with a bin size of 0.2 kpc.

For each bin, we run the WT and determine the locations and significance of each overdensity. By plotting each detected peak on the kinematic plane, we can track the evolution of the moving groups throughout the Galactic disk. This is shown in Figure~\ref{fig:worms} (some extraneous detections not associated with a continuous stream have been removed). Each dot is a detected peak colored by its Galactocentric radius. The moving groups in the SN are shown as square markers and are labelled. Here we can clearly see that many of the detected moving groups extend $\gtrsim$1 kpc radially throughout the Galactic disk. The tracked groups with their radial extents are listed in Table~\ref{tab:rextent}. We also note that there are \rev{five}{four} groups with very limited radial extent ($<0.5$ kpc): Coma Berenices, HR1614, Pleiades, \rev{Boblyev16-23, and a previously uncharacterized group, (Group 21)}{and Zhao09-9}. A discussion of the differences between these and the radially extended groups is included in Section \ref{sec:disc}.

\subsubsection{Shapes of Moving Groups in the Kinematic Plane} \label{sec:contours}

In addition to simply detecting the peaks of overdensities, the WT evaluates the wavelet coefficients across the entire image (shown as green and purple shaded regions in Figure~\ref{fig:results}b and c). We can then look at the shape of the moving groups in kinematic space by plotting contours of constant wavelet coefficient. We have also performed this analysis as a function of radius and the results are shown in Figure~\ref{fig:contours}. Note that while the contour levels are consistent across radius within a single group (e.g. all contours for the Hercules group are 10\% of the maximum wavelet coefficient at each radius), the contour levels vary from group to group (e.g. the contours for Sirius are at the 40\% level whereas the contours for Hyades are at the 90\% level). This allows for optimum visualization of groups with different wavelet coefficient values, however this means that the relative size between groups in this figure does not have meaning. The main purpose of this figure is to show how the kinematics of individual groups changes with radius.

For example, as we progress towards the Galactic center, we can see that Hercules covers a larger portion of the kinematic plane. Therefore, at smaller radii, the percentage of stars in the Hercules stream increases. Converesely, as we progress towards the outer disk, Hercules tends to disappear. We see a similar but inverse trend with Sirius. At smaller Galactocentric radii, the contours around Sirius shrink and eventually vanish, but as we progress past the SN and beyond into the outer disk, Sirius grows to cover a significant portion of the kinematic plane. %This is discussed further below.

\section{Discussion} \label{sec:disc}

\begin{figure}
    \centering
    \includegraphics[width=\columnwidth]{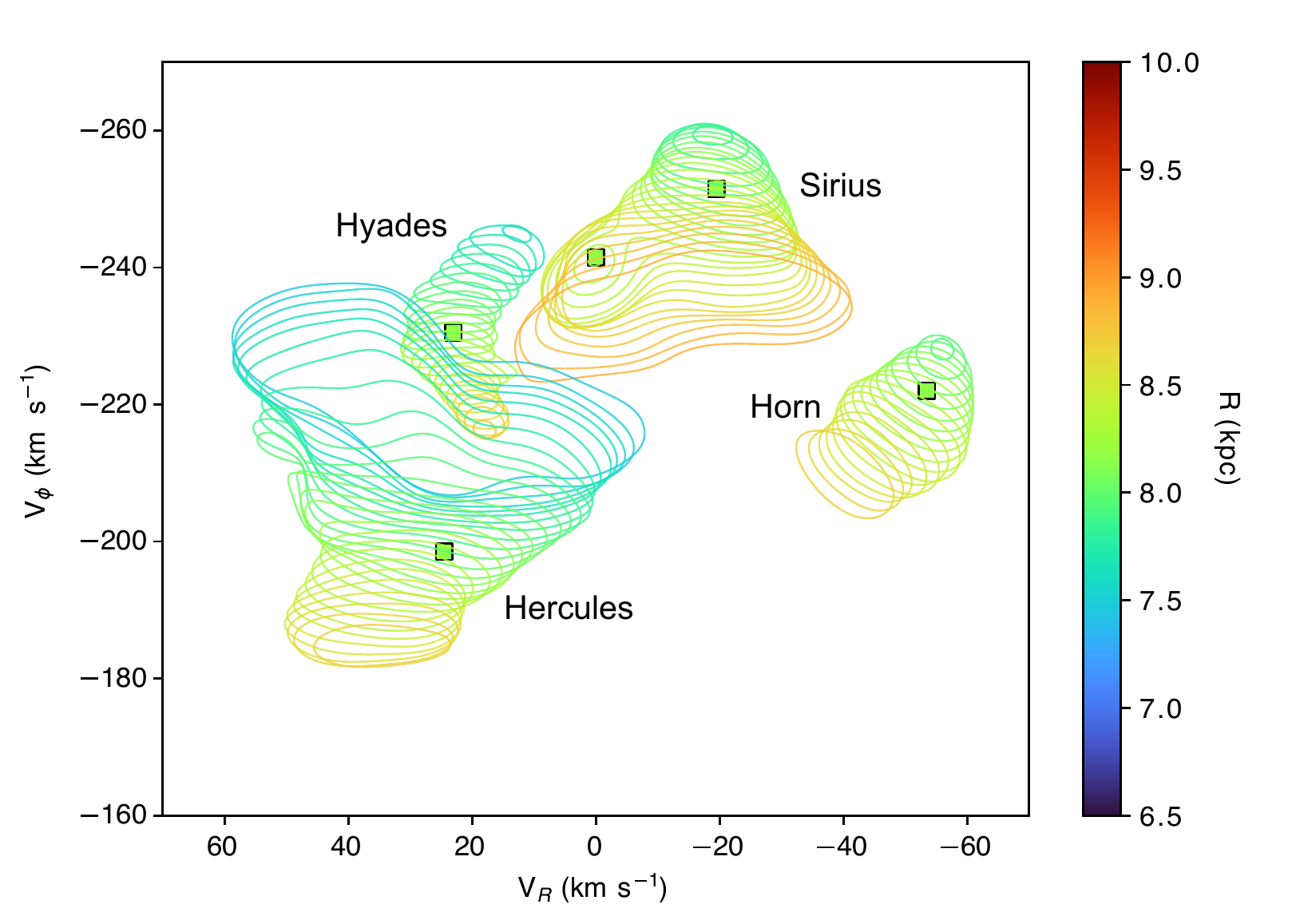}
    \caption{As in Fig \ref{fig:worms}, except each moving group is designated by contours of constant wavelet coefficient. Note that the axis limits are different from Figure~\ref{fig:worms}. This allows us to see how the shapes of the groups vary as a function of radius. Note only four groups are displayed here to maintain clarity.}
    \label{fig:contours}
\end{figure}

Our WT code, \textit{MGwave}, performs 2D wavelet transformations with the goal of detecting statistically significant circular overdensities and underdensities at varying scales. This is distinct from many recent WT analyses of the SN kinematic plane.
% Moreover, our method is distinct from many recent WT phase space analyses.
\citetalias{ramos18} don't include a minimum star count cutoff and thus detect many more fringe overdensities that we consider not significant against Poisson noise. \citet{yang21} use a bivariate WT to detect features in the $V$ vs $\sqrt{U^2+2V^2}$ space and they utilize a Gaussian mixture model with Monte Carlo sampling to generate a smooth background distribution to compare against. \citet{bernet22} explore the $V_R-V_\phi$ plane by performing a one-dimensional WT on slices in $V_R$. By linking peaks in the 1D WT with neighboring $V_R$ bins, they detect arches in the kinematic plane analogous to those found in \citetalias{ramos18}.
% This is in contrast to our \textit{MGwave} code which performs a 2D WT with the goal of detecting circular overdensities at varying scales.

% \begin{figure}
%     \centering
%     \includegraphics[width=\columnwidth]{figs/phi_Herc_contours_clean.pdf}
%     \caption{The $V_R$-$V_\phi$ kinematic plane with contours of constant wavelet coefficient shown at varying azimuthal values around the MW disk. $\phi=0^\circ$ corresponds to the location of the sun, and negative $\phi$ is in the direction of rotation while positive $\phi$ is counter to the direction of rotation. Here the contours for Hercules are shown and they become larger and merge with the main mode as we move to positive $\phi$, i.e. towards the minor axis of the MW bar. Other significant SN moving groups are labeled for reference.}
%     \label{fig:phicontours}
% \end{figure}

\begin{figure*}
    \centering
    \includegraphics[width=\textwidth]{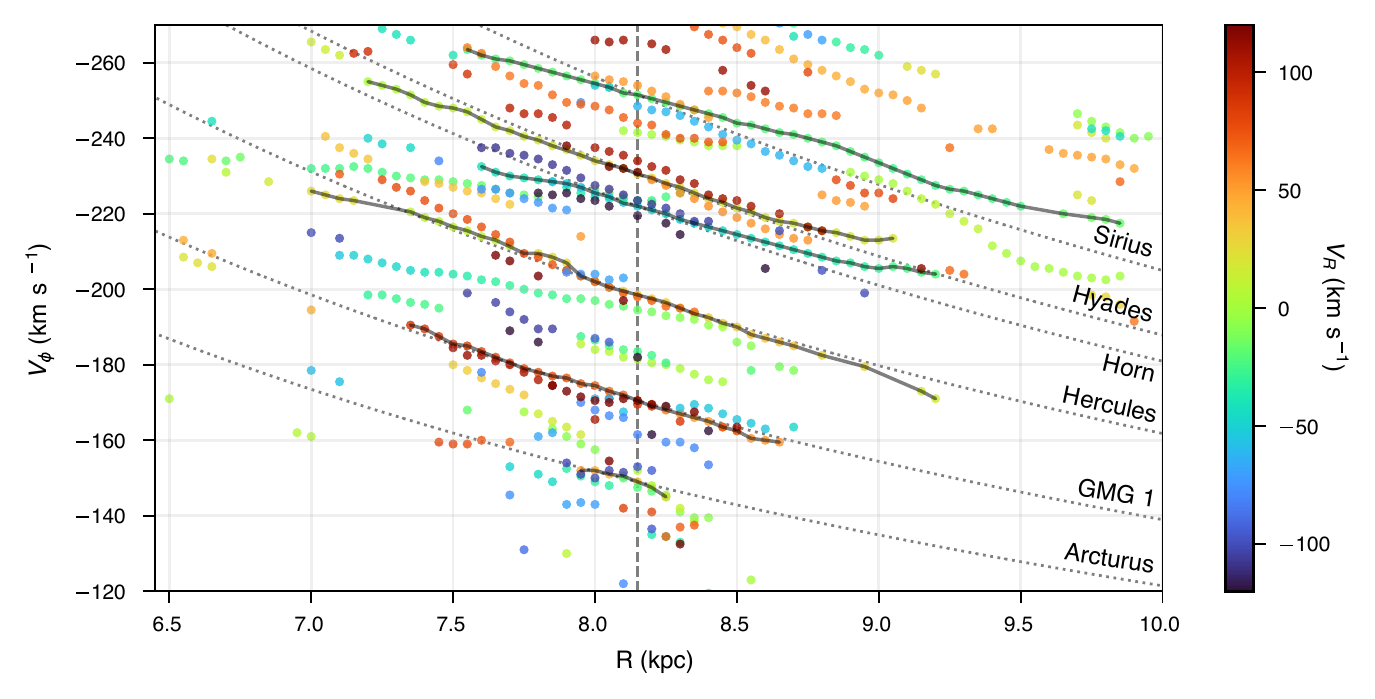}
    \caption{The variation of $V_\phi$ with $R$. Each point is a detected overdensity in the kinematic plane colored by $V_R$. The solar neighborhood is shown with a vertical dashed line at 8.15 kpc. The continuous streams associated with Sirius, Hyades, the Horn, Hercules, GMG 1, and Arcturus are shown as connected points and curves of constant angular momentum are shown as dashed lines. \rev{}{Note that only those points with $P_\mathrm{MC}\geq0.8$ are shown.}
    }
    \label{fig:vphir}
\end{figure*}

% \subsection{Across the MW Disk}

While previous works have analyzed moving groups through radius (e.g. \citetalias{ramos18}; \citealt{antoja18,fragkoudi19,bernet22}), they have focused on the variation in the locations of the peak overdensities (e.g. Figures~\ref{fig:worms}, \ref{fig:vphir}). Our Figure~\ref{fig:contours} shows that the WT can provide much richer information than simply the location of the extrema. The contours of these groups and how they evolve with radius and azimuth can be informative on the properties of the non-axisymmetric features of the Galactic disc.
%break the degeneracies in bar pattern speed and help determine the properties of our Galaxy.
Because many of these groups are so extended in radius, we know that they are not local, transient structures, but large-scale features of the MW disk. Their extent indicates that these moving groups are likely formed through the gravitational effects of the MW's non-axisymmetric features.

%As discussed in the introduction, 
The MW's bar and spiral arms and their associated resonances have long been used to explain the origin of moving groups. \rev{}{The specific resonances that are able to form the groups depend on the bar model (e.g. Hercules can be formed by the outer Lindblad resonance of a short bar or by the corotation resonance of a longer bar). However,} recent works seem to indicate that a long bar with pattern speed of $\sim$40 \kmskpc\ is consistent with both direct observations \citep{clarke19,sanders19} and can explain
%implications for 
many of the moving groups that we detect in the SN \citep{monari19,donghia20,trick21}. 
\citet{donghia20} proposed a model with a bar of length 4.5 kpc and pattern speed of 40 \kmskpc and showed that Hercules is reproduced by stars at the corotation resonance with the bar. In this scenario Hercules' stars are librating around the bar's Lagrange points L4/L5 thus leading to a stream of stars with coherent velocity (slower than the sun) in the SN (see their Figure~4).
% \rev{}{Previous works have also suggested that Hercules could be formed through the Outer Lindblad Resonance of a shorter bar. However, while both of these models are degenerate in the solar neighborhood, by looking throughout the Galactic disk, we can break these degeneracies and constrain properties of the MW's bar.} 
As shown in our Figures~\ref{fig:worms} and \ref{fig:contours}, Hercules is extended in radius around the SN. Moreover, Figure~\ref{fig:contours} shows that Hercules grows to cover a significant portion of the kinematic plane for $R < R_\odot$. In the models of \citet{donghia20}, the bar’s corotation radius is around 6 kpc, so if the stars of Hercules are formed through trapping at corotation, we would expect Hercules to become more significant at smaller radii, consistent with the data.

\begin{figure}
	\centering
	\includegraphics[width=\columnwidth]{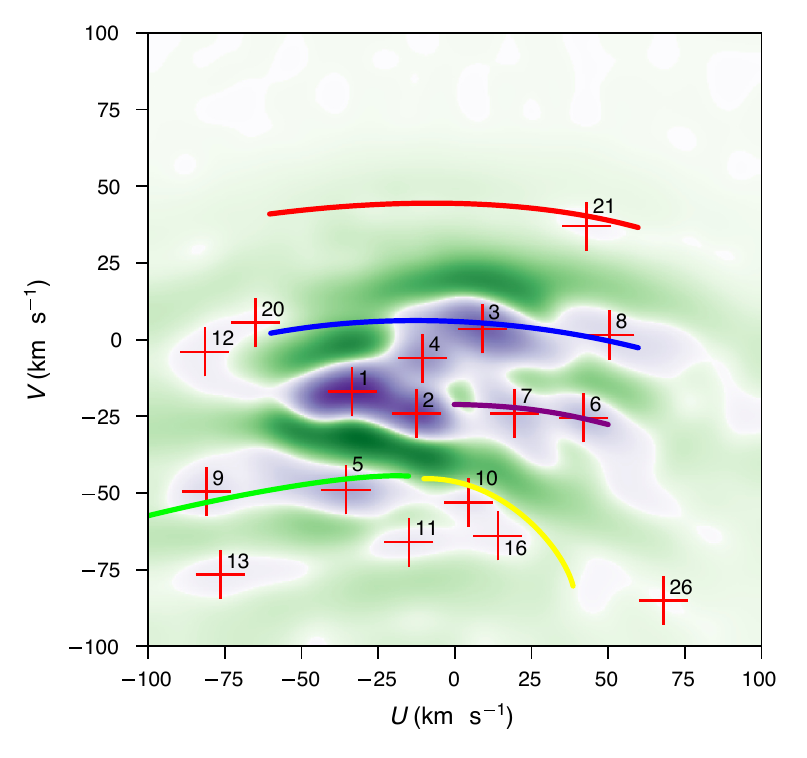}
	\caption{The $j=4$ wavelet transformed image of the solar neighborhood in $U,V,W$ coordinates. Coordinates were transformed from \gaia\ DR3 source data to the Astropy default Galactic coordinate system (in the cartesian representation). The overdensities are numbered as in Figure~\ref{fig:numbered}. Overlaid are the bar resonance locations from \citet{monari19} (c.f. their Figure~9). ``The green and yellow lines correspond to the corotation, the red line to the 2:1 (OLR) resonance for the m = 2 mode, ... the blue line to the 4:1 resonance of the m = 4 mode, and the purple line to the 6:1 resonance of the m = 6 mode.''}
	\label{fig:monari}
\end{figure}

The model of a long bar presented in \citet{monari19} also shows that five regions of in the kinematic plane correspond to resonances with the bar. 
%in their bar potential (with a pattern speed of 39 \kmskpc). They find that 
\rev{}{To compare with this work, we performed the WT on the \gaia\ DR3 data transformed into $U,V,W$ coordinates\footnote{We used the default Galactic coordinate frame in the Astropy Python module \citep{astropy:2022}}. Figure~\ref{fig:monari} shows this WT image and the corresponding overdensities numbered by their corresponding group in Table~\ref{tab:dr3}. The colored lines show the locations of the resonances from the long bar model of \citet{monari19}: red, blue, and purple correspond to the 2:1 (OLR), 4:1 (outer ultra-harmonic resonance, OUH), and 6:1 resonances, while the green and yellow lines mark the corotation resonance.}
In addition to Hercules being stars at corotation with the bar, the authors find that the Hat aligns with OLR, Sirius with the \rev{Outer Ultra-Harmonic resonance (OUH)}{OUH}, and the Horn with the 6:1 resonances. All of these groups are shown in our Figure~\ref{fig:worms} and are \rev{significantly }{}still prominent across \rev{the }{}Galactocentric radius. Moreover, Figure~\ref{fig:contours} shows Sirius becoming more prominent at larger radii (opposite of Hercules). For a bar with a pattern speed of $\sim$40 \kmskpc, the location of the OUH is at 8.5-9 kpc \citep{donghia20}. Therefore, we expect more stars comprising Sirius as we look towards the outer Galactic disk, which is shown in the data.
% Therefore, the data show that these groups are most likely formed through resonances due to their large-scale structure and extent.

There are several other groups that we detect with significant radial extent, many of which are also identified being in resonance with the bar \citep{monari19} \rev{}{shown in Figure~\ref{fig:monari}}. Antoja12-16, Antoja12-GCSIII-13, $\gamma$Leo, \rev{new group DR3G-21}{Zhao09-9}, and possibly Antoja12-12 fall on the OUH resonance along with Sirius. Dehnen98-6 aligns well with the 6:1 resonance along with the Horn. Finally, $\epsilon$Ind and Hercules and Liang17-9 are all at corotation.

This leaves four groups with radial extent greater than 0.5 kpc unaccounted for: the Hyades, GMG 1, GMG 3, and Kushniruk17-J4-19.
%Its unique bimodal slope in $V_R$ (discussed above) perhaps alludes to the fact that it has been formed through an alternate pathway. 
While Hyades doesn't seem to have formed through any known bar resonance, works focusing on the kinematic signatures of spiral arm resonances have been able to reproduce Hyades along with several other features of the SN kinematic plane (\citealt{michtchenko18,barros20}; including low $V_\phi$ features like GMG 1,3, and Kushniruk17-J4-19). However, these models predict that the moving groups are significantly extended in $\phi$, and less extended in $R$.
% As shown in Figure~\ref{fig:phicontours}, Hercules diminishes quite quickly in the negative $\phi$ direction, but 
Further work will be required to constrain the groups in $\phi$ to test this theory. % As \gaia\ data extends further in $\phi$ we will be able to test this theory.
Additionally, our detection of Hyades throughout a large range of Galactocentric radii could be simply the detection of the main mode in each neighborhood. We expect a smooth evolution of $V_\phi$ across radius with stars being mostly on circular orbits.
%and an average $V_R$ of 0 \kms. 
Therefore, while the main mode might be identified as Hyades locally, at different radii, the detected peak could simply be the bulk motion of the disk. This would also explain why it is unique in its double slope in $V_R$ in Figure~\ref{fig:results} (discussed further below).

There are also five groups detected that have small radial extent ($<0.5$ kpc): Coma Berenices, the Pleiades, HR1614, Boblyev16-23, and DR3G-21 (which was briefly discussed above). It has been shown previously that Coma Berenices and the Pleiades are open clusters %localized in physical space as well as in phase space 
\citep[e.g.][]{odenkirchen98,tang18,heyl22}. Figure~\ref{fig:worms} and Table \ref{tab:rextent} corroborates this result by showing that these objects are detected only locally within the SN. While HR1614 has long been considered an open cluster \citep[e.g.][]{feltzing00,desilva07} recent works suggest that its metallicity spread matches that of the MW disk population \citep{kushniruk20}. Further investigation is required to unravel the true origin of HR1614.
% In \citet{kushniruk17}, they discuss the properties of Kushniruk17-J3-13 which is located at the same phase space position as Kushniruk17-J4-2. Their group has 201 thin and thick disk stars with a metallicity distribution consistent with the entire solar neighborhood sample (from \gaia\ DR1 and RAVE). They state that this group could be an elongation of nearby Sirius or $\gamma$Leo as their properties are similar. While Figure~\ref{fig:worms} and Table \ref{tab:rextent} indicate that it is simply a local structure (due to its small radial extent), further spectroscopic follow up is required to determine whether Kushniruk-J4-2 is a dynamical moving group or an open cluster.

%In addition to studying the length of the groups, much can be learned from the variation in $V_\phi$ and $V_R$ with $R$. Previous works have explored the variation of $V_\phi$ with $R$ in order to explore the evolution of the angular momentum of each moving group across the disk (e.g. \citetalias{ramos18}; \citealt{antoja18,fragkoudi19,bernet22}). 
We apply our \textit{MGwave} code to \gaia\ DR3.
Figure~\ref{fig:vphir} shows 
the azimuthal velocity of the known moving groups of the SN  displayed as a function of Galactocentric radius. \rev{e}{E}ach moving groups is colored by radial velocity.  \citet{bernet22} showed the same plot but using a 1D WT technique applied to the previous Gaia data release (eDR3). The authors found that all major groups deviate from the predicted $V_\phi\propto R^{-1}$. Note that our results obtained with DR3 seem to confirm a deviation from the constant angular momentum curve (dashed line) for most of the known moving groups, with the exception of GMG 1. This general outcome is not surprising as the constant angular momentum curve is expected for small radial oscillations of the stars, within the epicycle approximation. Therefore, a deviation is expected for highly eccentric stars.
%However, their simulations also deviated from the naive constant angular momentum prediction as well indicating that even if a group is formed purely through resonances, it may deviate from this $R^{-1}$ dependence. 
Additionally, even with the improved data of \gaia\ DR3, we are unable to trace our groups much inwards of $R\sim7$~kpc while \citet{bernet22} find groups extending down to 5~kpc. This discrepancy could be due to the difference in the wavelet method (searching for arches vs. search for circular features). However it is also clear that the data become less accurate at these radii. While the WT is able to detect significant overdensities even at these small radii, many of them have small $P_\mathrm{MC}$ values indicating that they are not robust detections against the \gaia\ errors (see Section~\ref{sec:mcs}).

Our Figure~\ref{fig:worms} shows that there is also a significant variation in $V_R$ with $R$. Notably that most groups have a shift in $V_R$ as they move in radius, however the direction of this shift (the slope of the connected points in Figure~\ref{fig:worms}) can be positive, negative, or both. Four groups have strong positive slopes (e.g. Liang17-9, Dehnen98-6, the Horn, and \rev{portions}{the majority} of Hercules) in which they move to larger $V_R$ at higher radii, and three have strong negative slopes (GMG 3, GMG 1, and Kushniruk17-J4-19) with smaller $V_R$ at higher radii. Several other groups have slight slopes in either direction, or multiple slopes at different radii. Notably, Hyades moves to larger $V_R$ until it reaches the solar neighborhood at which point it decreases again, and \rev{the Horn moves to constant $V_\phi$ at its innermost extent}{Hercules, Sirius, and Dehnhn98-6 exhibit breaks or strong variations in the slope throughout radius}.

For the positive slope groups, the inner portion (smaller $R$) has an inward velocity relative to the centroid, while the outer portion (larger $R$) has an outward velocity. This will inevitably lead to the group spreading out and possibly breaking apart. Consequently, negative slope groups exhibit the opposite trend and therefore are condensing. These two different behaviors of groups could possibly indicate environmental effects operating at different radii like tidal effects, but further analysis of the data in comparison with simulations is required to fully explore the possible causes of these slopes.

\begin{figure}
	\centering
	\includegraphics[width=\columnwidth]{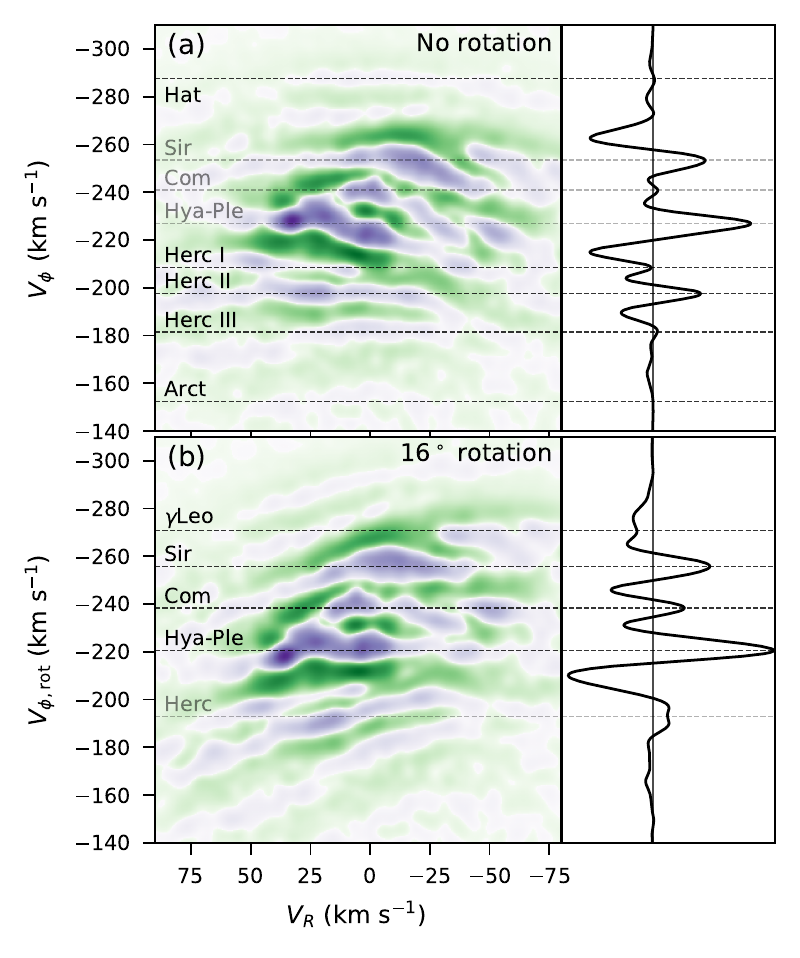}
	\caption{The wavelet plane and its histogram along $V_\phi$ for $j=3$, structures of size 4-8~km~s$^{-1}$. In the top panels, the wavelet plane is identical to that shown in Figure~\ref{fig:sn_scales}, however on the bottom panels, the plane has been rotated by 16$^\circ$ counter-clockwise. As done in \citet{antoja08}, this is to better align the Hyades-Pleiades, Coma Berenices, and Sirius overdensities with the $x$-axis. However, as shown in the top panel, there are several structures (Hercules, the Hat, and Arcturus) which do not match this distinctive rotation. The line plots on the right show the sums of wavelet coefficients across all $V_R$ as a function of $V_\phi$. These panels show that the Hyades-Pleiades, Coma Berenices, and Sirius overdensities are stronger and more prominent after a 16$^\circ$ rotation of the plane, whereas Hercules, the Hat, and Arcturus are stronger without the rotation.
	}
	\label{fig:hist}
\end{figure}

%\subsection{Kinematic Branches} \label{sec:branches}

Looking at the larger structure of the WT images, previous works \citep[e.g.][]{skuljan99,antoja08} have noted several distinct kinematic branches visible in the $V_R-V_\phi$ plane. \rev{}{Note that these features have been explored at smaller scales that those discussed throughout most of this paper. In the following paragraphs we will be referencing our results from the WT with scale $j=3$.} \citet{antoja08} found that these branches are inclined at an angle of $\sim$16$^\circ$ and the four most prominent are aligned with the Hercules, Hyades/Pleiades, Coma Berenices, and Sirius groups. These branches are still clearly visible in our data (see Figure~\ref{fig:results}c), however thanks to \gaia's immense volume of data, we can now view these structures across larger ranges of $V_R$. \citet{gaia18} and \citetalias{ramos18} extended these branches into arches, most of which follow constant kinetic energy. In contrast to the uniformly inclined branches found in \citet{antoja08}, \gaia's increased data has elongated and straightened out many of these structures. However, as discussed in \citetalias{ramos18}, several of the arches are still inclined to one side; notably their A5 and A7 corresponding to Hyades/Pleiades and Coma Berenices.
% However, the four main branches they originally detected are still clearly visible in a histogram of the $V_\phi$ values of the wavelet transformed kinematic plane, shown in Figure~\ref{fig:hist}.
\rev{}{Several models have shown that these arches of constant kinetic energy can be formed through phase mixing (\citealt{minchev09,gomez12}; \citetalias{ramos18}) which could play a role in the formation of the moving groups as well. However further investigation is required to constrain this paradigm.}

In this study we have returned to the method of \citet{antoja08} of summing the wavelet transformed image along $V_R$ to obtain a histogram as a function of $V_\phi$. We have explored this histogram both with and without the 16$^\circ$ rotation that was performed in \citet{antoja08}. These results can be seen in Figure~\ref{fig:hist} (Panels (a) show the results without rotation and Panels (b) include the 16$^\circ$ rotation). With the %dramatically% 
increased volume of data provided by \gaia, we can see that, while the Hyades/Pleiades, Coma Berenices, and Sirius branches do still appear inclined (and we see $\gamma$Leo appear in the rotated histogram as well), there are several other structures that do not follow this trend. Most dramatically, we see three strong peaks in the non-rotated histogram corresponding to various components of Hercules. By looking at the wavelet plane, the horizontal alignment of these branches is clearly visible, while in the rotated plane (bottom panels) Hercules becomes muddled. We also see slight peaks in the non-rotated histogram corresponding to the Hat at very high $V_\phi$, and Arcturus at very low $V_\phi$.
While some of the these tilted features have been reproduced in past simulations \citep[e.g.][]{antoja09,hunt18,barros20}, further modeling is required to determine their source specifically in the context of a long, slow bar.

% \citet{antoja09} explored the combined effects of a bar and spiral arms to create asymmetric features in the SN kinematic plane. By including bars at various pattern speeds with and without spiral arms, they can reproduce horizontal features (as we see in Hercules, Arcturus, and the Hat) as well as asymmetric overdensities (such as our Hyades/Pleiades, Coma, and Sirius). More recent simulations exploring the effect of spiral arms \citep{hunt18,barros20} also reproduce both tilted and horizontal features in phase space. While further modeling of the possible interactions between the bar and spiral arms in the MW is required, the general agreement between Figure~\ref{fig:hist} and these studies suggests that both the bar and spiral arms play a large role in shaping our solar neighborhood.

\section{Conclusions} \label{sec:conclusions}

% As the kinematic data of the stars in our Milky Way continue to improve, we will be able to utilize techniques like the wavelet transformation discussed here to quickly and easily analyze the complex structures of our Galaxy. By learning more about the intricacies of the Galactic disk, we can better determine the total mass, gravitational potential, information about the bar and spiral arms, as well as the satellite interaction history. As our data sets become larger and larger, we will need robust methods like these to continue to uncover the mysteries of our Galaxy.

The wavelet transform is an invaluable tool for precise, quantitative analysis of images. Our new code, \textit{MGwave}, is an open-source Python module for performing wavelet transformations on 2D images while detecting extrema and determining their significance. Additionally, we have implemented Monte Carlo sampling to propagate errors and uncertainties through to the wavelet extrema detections. \textit{MGwave} is able to reproduce the findings of \citetalias{ramos18} (using \gaia\ DR2 data) and improves upon previous codes by detecting underdensities in addition to overdensities and implementing a minimum $n$ cutoff in the significance calculation.

% We performed the WT on \gaia\ DR3 data to detect moving groups in the kinematic plane ($V_R-V_\phi$) of the solar neighborhood (Figure~\ref{fig:results}). With the improved data, we have detected five new, statistically significant candidate moving groups. We also analyzed the kinematic branches of the SN kinematic plane by looking at the alignment of the overdensities within the main groups (Sirius, Coma Berenices, Hyades, and Pleiades) compared with that of the fringe groups (the Hat, Hercules, and Arcturus). The immense improvement in data since \citet{antoja08} (due to \gaia) has allowed us to show that while the main groups do still appear to be inclined at $\sim$16$^\circ$, the fringe groups do not follow this alignment. This could indicate a fundamental difference between these two categories of moving groups.

We performed the WT on \gaia\ DR3 data to detect moving groups in the kinematic plane ($V_R-V_\phi$) of the solar neighborhood (Figure~\ref{fig:results}). With the improved data, we have several main conclusions:
\begin{itemize}
    \item We have detected \rev{four}{three} new, statistically significant candidate moving groups: \rev{one near Antoja12-GCSIII-13, one near GMG 5, and two more at very low $V_R$ with similar $V_\phi$ to Hercules}{one within Arcturus, and two in regions without much substructure at low $V_R$}.
    \item We have been able to perform the WT on different regions within the MW disk. Exploring the structure of the kinematic plane in sections of the disk ranging in Galactocentric radius from 6.5 to 10~kpc, we find that the majority of the moving groups detected within the SN are radially extended (Figure~\ref{fig:worms}). The elongation of these groups indicate that they are dynamical structures possibly outcome  by the effects of resonances of the MW's non-axisymmetric features.
    \item By mapping contours in wavelet space, we can track the variation in the kinematic shape of these groups through radius (Figure~\ref{fig:contours}). We find Hercules becoming more prominent towards the galactic center, in agreement with the models of \citet{donghia20} \rev{predicting}{that predicted} that Hercules is comprised of stars at corotation with the bar.
    \item Mapping WT contours also reveals an opposite trend for Sirius, it gets more prominent towards the outer disc. This \rev{seems to be}{is} consistent with Sirius being in resonance with the OUH located outside the solar radius \citep{monari19}.
    % \item We have used the WT to track groups in azimuth (Figure~\ref{fig:phicontours}). Hercules becomes much more prominent in kinematic space as we move against the direction of rotation of the disk, i.e. towards the semi-minor axis of the MW bar. This was also predicted by \citet{donghia20} if the Hercules stars are librating around the Lagrange points located along the semi-minor axis L4/L5.
\end{itemize}

\gaia\ DR3 has greatly expanded our view of the MW. By looking at the kinematics of moving groups throughout a significant portion of the disk, we can unravel many of the mysteries of the MW's non-axisymmetric features and their associated resonances.

\section*{Acknowledgements}
The authors thank the anonymous referee for their constructive comments on the manuscript. The authors also thank Eric Slezak for useful discussions on the implementation of the wavelet significance calculations. This work made use of Astropy:\footnote{http://www.astropy.org} a community-developed core Python package and an ecosystem of tools and resources for astronomy \citep{astropy:2013, astropy:2018, astropy:2022}.

\section*{Data Availability}
The data underlying this article will be shared on reasonable request to the corresponding author.

\newpage
\bibliographystyle{mnras}
\bibliography{references}

\appendix

% \section{Derivation of Wavelet Equation}
% \label{a:psidef}
% We begin with the definitions of wavelet function, $\psi$, and the scaling function, $\phi$, in terms of the filters $h$ and $g$:
% \begin{align*}
%     \frac{1}{4}\phi\left(\frac{x}{2},\frac{y}{2}\right) &= \sum_{\ell,m} h(\ell,m)\phi(x-\ell,y-m) \\
%     \frac{1}{4}\psi\left(\frac{x}{2},\frac{y}{2}\right) &= \sum_{\ell,m} g(\ell,m)\phi(x-\ell,y-m)
% \end{align*}
% We are using the Starlet transform which defines $g=\delta - h$:
% \begin{align*}
%     \frac{1}{4}\psi\left(\frac{x}{2},\frac{y}{2}\right) &= \sum_{\ell,m} \big(\delta(\ell,m) - h(\ell,m)\big)\phi(x-\ell,y-m) \\
%     &= \phi(x,y) - \sum_{\ell,m} h(\ell,m)\phi(x-\ell,y-m) \\
%     &= \phi(x,y) - \frac{1}{4}\phi\left(\frac{x}{2},\frac{y}{2}\right)
% \end{align*}

\begin{table*}
    \centering
    \begin{tabular}{rrrlcrccc}
& $V_R$ & $V_\phi$ & Name & CL & $P_\mathrm{MC}$ & Wavelet & n & Stars \\\hline
 1 &   22.5 & -236.0 & Hyades             & 3 & 1.00 & 7.6596 & 394 & 130,982 \\
 2 &    1.5 & -228.5 & Pleiades           & 3 & 1.00 & 5.6223 & 438 & 162,864 \\
 3 &  -20.0 & -256.5 & Sirius             & 3 & 1.00 & 4.9166 & 376 & 87,694 \\
 4 &   -1.0 & -247.0 & Coma Berenices     & 3 & 1.00 & 3.0033 & 538 & 130,122 \\
 5 &   24.0 & -203.5 & Hercules II        & 3 & 1.00 & 2.0458 & 157 & 58,368 \\
 6 &  -54.0 & -227.5 & Dehnen98-14        & 3 & 1.00 & 1.5060 & 141 & 44,723 \\
 7 &  -31.0 & -228.5 & Dehnen98-6         & 3 & 1.00 & 1.2256 & 293 & 90,548 \\
 8 &  -64.0 & -253.5 & $\gamma$Leo        & 3 & 1.00 & 0.5601 & 68 & 20,072 \\
 9 &   70.5 & -203.0 & $\epsilon$Ind      & 3 & 1.00 & 0.5349 & 43 & 17,400 \\
10 &  -53.0 & -258.0 & Kushniruk17-J5-2*  & 3 & 0.93 & 0.4607 & 112 & 23,082 \\
11 &  -16.5 & -199.5 & Liang17-9          & 3 & 1.00 & 0.2771 & 109 & 38,112 \\
12 &   70.0 & -250.5 & Antoja12-GCSIII-13 & 3 & 1.00 & 0.1807 & 24 & 9,614 \\
13 &   64.0 & -239.0 & Dehnen98-11*       & 3 & 0.99 & 0.1570 & 61 & 17,261 \\
14 &   66.0 & -175.5 & GMG 1              & 3 & 1.00 & 0.1473 & 20 & 9,255 \\
15 &   -7.0 & -187.5 & GMG 2              & 3 & 1.00 & 0.1293 & 102 & 31,186 \\
16 & \textbf{-68.0} & \textbf{-210.0} & \textbf{Unknown} & \textbf{3} & \textbf{0.47} & \textbf{0.1248} & \textbf{54} & \textbf{22,073} \\
17 & -108.5 & -229.0 & Antoja12-12        & 3 & 1.00 & 0.1208 & 11 & 3,764 \\
18 &  106.5 & -239.5 & Antoja12-16        & 3 & 1.00 & 0.1111 & 6 & 2,233 \\
19 &    2.5 & -186.0 & Kushniruk17-J4-19* & 3 & 0.60 & 0.1077 & 76 & 33,118 \\
20 &   47.0 & -178.0 & Arifyanto05        & 3 & 0.86 & 0.0898 & 34 & 14,832 \\
21 &   88.5 & -174.5 & GMG 3              & 3 & 0.98 & 0.0708 & 6 & 3,953 \\
22 &   36.0 & -153.5 & $\eta$Cep          & 3 & 0.93 & 0.0602 & 12 & 5,041 \\
23 &  -51.0 & -291.0 & GMG 4              & 3 & 1.00 & 0.0529 & 6 & 1,345 \\
24 &  -28.5 & -189.0 & HR1614*            & 1 & 0.81 & 0.0314 & 59 & 20,115 \\
25 &  -26.0 & -150.0 & Antoja12-17        & 3 & 1.00 & 0.0311 & 7 & 3,087 \\
26 &   48.0 & -259.5 & Zhao09-9*        & 0 & 1.00 & 0.0292 & 71 & 14,006 \\
27 &  -56.0 & -176.0 & GMG 5              & 3 & 1.00 & 0.0276 & 14 & 5,037 \\
28 &  106.0 & -272.5 & GMG 6              & 3 & 0.99 & 0.0254 & 1 & 318 \\
29 & -134.0 & -225.5 & GMG 7              & 3 & 1.00 & 0.0173 & 1 & 535 \\
30 & \textbf{2.5} & \textbf{-157.5} & \textbf{Unknown} & \textbf{2} & \textbf{0.98} & \textbf{0.0170} & \textbf{14} & \textbf{6,182} \\
31 & \textbf{-14.0} & \textbf{-153.0} & \textbf{Unknown} & \textbf{1} & \textbf{0.37} & \textbf{0.0159} & \textbf{17} & \textbf{4,305} \\
32 &  129.0 & -237.5 & GMG 8              & 3 & 1.00 & 0.0159 & 1 & 496 \\
33 & \textbf{112.5} & \textbf{-155.0} & \textbf{Unknown} & \textbf{3} & \textbf{0.68} & \textbf{0.0123} & \textbf{3} & \textbf{1,078} \\
34 & \textbf{-88.0} & \textbf{-233.0} & \textbf{Unknown} & \textbf{0} & \textbf{1.00} & \textbf{0.0120} & \textbf{40} & \textbf{10,436} \\
35 &   73.0 & -282.0 & GMG 11             & 2 & 1.00 & 0.0112 & 4 & 770 \\
36 & -108.0 & -152.0 & GMG 12             & 3 & 0.99 & 0.0106 & 1 & 345 \\
37 &  125.5 & -175.0 & GMG 14             & 3 & 1.00 & 0.0104 & 3 & 654 \\
38 &  -78.0 & -130.5 & GMG 13             & 2 & 0.97 & 0.0101 & 4 & 500 \\
39 & \textbf{83.5} & \textbf{-143.5} & \textbf{Unknown} & \textbf{2} & \textbf{0.97} & \textbf{0.0095} & \textbf{2} & \textbf{1,285} \\
40 & \textbf{-79.0} & \textbf{-166.5} & \textbf{Unknown} & \textbf{1} & \textbf{0.99} & \textbf{0.0092} & \textbf{8} & \textbf{1,671} \\
41 &   -1.0 & -120.0 & GMG 17             & 1 & 0.97 & 0.0090 & 6 & 709 \\
42 &  -24.0 &  -91.0 & GMG 16             & 3 & 1.00 & 0.0085 & 1 & 219 \\
43 & \textbf{71.0} & \textbf{-142.5} & \textbf{Unknown} & \textbf{2} & \textbf{0.49} & \textbf{0.0082} & \textbf{3} & \textbf{1,431} \\
44 & \textbf{-59.0} & \textbf{-138.0} & \textbf{Unknown} & \textbf{3} & \textbf{0.44} & \textbf{0.0070} & \textbf{1} & \textbf{1,019} \\
45 &  -39.0 & -139.0 & Antoja12-19*       & 1 & 0.90 & 0.0067 & 6 & 1,568 \\
46 &   13.5 &  -82.0 & GMG 19             & 3 & 0.85 & 0.0067 & 1 & 158 \\
47 &  122.0 & -202.5 & GMG 20             & 1 & 0.96 & 0.0065 & 5 & 837 \\
48 & \textbf{-96.5} & \textbf{-158.5} & \textbf{Unknown} & \textbf{1} & \textbf{0.78} & \textbf{0.0063} & \textbf{4} & \textbf{636} \\
49 & \textbf{-66.5} & \textbf{-108.0} & \textbf{Unknown} & \textbf{3} & \textbf{0.97} & \textbf{0.0061} & \textbf{1} & \textbf{298} \\
50 &  -42.5 & -119.0 & GMG 22             & 1 & 1.00 & 0.0040 & 2 & 591 \\
51 & \textbf{125.0} & \textbf{-146.5} & \textbf{Unknown} & \textbf{1} & \textbf{0.49} & \textbf{0.0037} & \textbf{2} & \textbf{472} \\
52 & \textbf{-26.5} & \textbf{-107.0} & \textbf{Unknown} & \textbf{1} & \textbf{0.83} & \textbf{0.0033} & \textbf{1} & \textbf{384} \\
53 & \textbf{137.5} & \textbf{-164.5} & \textbf{Unknown} & \textbf{1} & \textbf{0.66} & \textbf{0.0028} & \textbf{1} & \textbf{381} \\
54 & \textbf{-98.5} & \textbf{-135.0} & \textbf{Unknown} & \textbf{1} & \textbf{0.88} & \textbf{0.0028} & \textbf{1} & \textbf{304} \\
55 & \textbf{136.5} & \textbf{-138.0} & \textbf{Unknown} & \textbf{1} & \textbf{0.93} & \textbf{0.0025} & \textbf{2} & \textbf{248} \\
56 & \textbf{-127.5} & \textbf{-184.5} & \textbf{Unknown} & \textbf{0} & \textbf{0.89} & \textbf{0.0022} & \textbf{2} & \textbf{486} \\
57 &  -70.5 & -174.0 & Antoja12-15*       & 0 & 0.42 & 0.0022 & 6 & 3,081 \\
58 & \textbf{109.0} & \textbf{-118.5} & \textbf{Unknown} & \textbf{1} & \textbf{0.60} & \textbf{0.0021} & \textbf{1} & \textbf{203} \\
59 & \textbf{140.0} & \textbf{-83.0} & \textbf{Unknown} & \textbf{1} & \textbf{0.96} & \textbf{0.0020} & \textbf{1} & \textbf{42}
    \end{tabular}
    \caption{Moving groups detected using our new wavelet transform on \gaia\ DR2 data. Compare with those found in \citetalias{ramos18}. Groups marked with an asterisk (*) are those that have been previously discovered but were not present in the wavelet analysis of \citetalias{ramos18}. Bold lines are groups newly discovered in this work. Columns 5-9 list the output of our analysis: CL denotes the confidence level that a given group is not due to Poisson noise (see Section \ref{sec:significance}); $P_\mathrm{MC}$ gives the percentage of Monte Carlo simulations in which the peak appeared when varying the stellar velocities within Gaia errors (see Section \ref{sec:mcs}); Wavelet gives the magnitude of the wavelet coefficient at the peak; n lists the number of stars used in computing the wavelet coefficient; Stars lists the number of stars in a region of kinematic space around the peak corresponding to the scale of the wavelet transformation performed (in this case within a circle of radius 16~km~s$^{-1}$).}
    \label{tab:dr2a}
\end{table*}

\begin{table*}
\begin{tabular}{rrrlcrccc}
60 & \textbf{50.5} & \textbf{-106.0} & \textbf{Unknown} & \textbf{1} & \textbf{0.54} & \textbf{0.0017} & \textbf{1} & \textbf{260} \\
61 & \textbf{-47.5} & \textbf{-88.0} & \textbf{Unknown} & \textbf{0} & \textbf{0.28} & \textbf{0.0016} & \textbf{2} & \textbf{180} \\
62 & \textbf{72.5} & \textbf{-78.0} & \textbf{Unknown} & \textbf{1} & \textbf{0.22} & \textbf{0.0015} & \textbf{1} & \textbf{113} \\
63 & -109.0 & -340.0 & GMG 27             & 1 & 0.98 & 0.0012 & 1 & 6 \\
64 &   47.0 & -342.0 & GMG 26             & 0 & 0.73 & 0.0012 & 2 & 13 \\
65 & \textbf{129.5} & \textbf{-125.0} & \textbf{Unknown} & \textbf{1} & \textbf{0.87} & \textbf{0.0011} & \textbf{1} & \textbf{200} \\
66 & \textbf{-114.5} & \textbf{-320.0} & \textbf{Unknown} & \textbf{1} & \textbf{0.96} & \textbf{0.0011} & \textbf{1} & \textbf{8} \\
67 & \textbf{-35.5} & \textbf{-74.0} & \textbf{Unknown} & \textbf{0} & \textbf{0.75} & \textbf{0.0011} & \textbf{2} & \textbf{139} \\
68 & \textbf{80.0} & \textbf{-116.5} & \textbf{Unknown} & \textbf{0} & \textbf{0.58} & \textbf{0.0010} & \textbf{3} & \textbf{306} \\
69 & \textbf{-25.0} & \textbf{-122.0} & \textbf{Unknown} & \textbf{0} & \textbf{0.88} & \textbf{0.0010} & \textbf{3} & \textbf{837} \\
70 & \textbf{-24.0} & \textbf{-326.5} & \textbf{Unknown} & \textbf{0} & \textbf{0.69} & \textbf{-0.0000} & \textbf{1} & \textbf{36} \\
71 & \textbf{-62.0} & \textbf{-153.0} & \textbf{Unknown} & \textbf{0} & \textbf{0.21} & \textbf{-0.0012} & \textbf{5} & \textbf{1,823} \\
72 & \textbf{-102.5} & \textbf{-183.5} & \textbf{Unknown} & \textbf{0} & \textbf{0.97} & \textbf{-0.0018} & \textbf{7} & \textbf{1,449} \\
73 & \textbf{-56.5} & \textbf{-161.5} & \textbf{Unknown} & \textbf{0} & \textbf{0.25} & \textbf{-0.0022} & \textbf{8} & \textbf{2,953} \\
74 & \textbf{-113.5} & \textbf{-199.0} & \textbf{Unknown} & \textbf{0} & \textbf{0.85} & \textbf{-0.0026} & \textbf{8} & \textbf{1,330}
\end{tabular}
\caption{\gaia\ DR2 moving group detections continued.}
\label{tab:dr2b}
\end{table*}

\label{lastpage}

\end{document}